\documentclass[review]{elsarticle}

\usepackage{lineno,hyperref}
\modulolinenumbers[5]

\journal{Journal of \LaTeX\ Templates}

\usepackage{graphicx}	% Including figure files
\usepackage{amsmath}	% Advanced maths commands

\usepackage{amssymb}	% Extra maths symbols
\usepackage{diagbox}
\usepackage[normalem]{ulem}
\usepackage[dvipsnames]{xcolor}
\usepackage{mathabx}

\begin{document}

\begin{frontmatter}

  \title{Mass classification of dark matter perturbers of stellar tidal streams}

%% Group authors per affiliation:
  \author{Francesco Montanari\fnref{fm}}
  \fntext[fm]{{E-mail: francesco.montanari@fmnt.info}}
  \author{Juan Garc\'ia-Bellido\fnref{jgb}}
  \fntext[jgb]{E-mail: juan.garciabellido@uam.es}
  \address{Instituto de F\'isica Te\'orica IFT-UAM/CSIC, Universidad
    Aut\'onoma de Madrid, Cantoblanco 28049 Madrid, Spain}

\begin{abstract}
  Stellar streams formed by tidal stripping of progenitors orbiting around the
  Milky Way are expected to be perturbed by encounters with dark matter
  subhalos. Recent studies have shown that they are an excellent proxy to
  infer properties of the perturbers, such as their mass. Here we present two
  different methodologies that make use of the fully non-Gaussian density
  distribution of stellar streams: a Bayesian model selection based on the
  probability density function (PDF) of stellar density, and a likelihood-free
  gradient boosting classifier. While the schemes do not assume a specific
  dark matter model, we are mainly interested in discerning the primordial
  black holes cold dark matter (PBH CDM) hypothesis form the standard particle
  dark matter one. Therefore, as an application we forecast model selection
  strength of evidence for cold dark matter clusters of masses $10^3$--$10^5
  M_{\Sun}$ and $10^5$--$10^9 M_{\Sun}$, based on a GD-1-like stellar stream
  and including realistic observational errors. Evidence for the smaller mass
  range, so far under-explored, is particularly interesting for PBH CDM. We
  expect weak to strong evidence for model selection based on the PDF
  analysis, depending on the fiducial model. Instead, the gradient boosting
  model is a highly efficient classifier (99\% accuracy) for all mass ranges
  here considered. As a further test of the robustness of the method, we reach
  similar conclusions when performing forecasts further dividing the largest
  mass range into $10^5$--$10^7 M_{\Sun}$ and $10^7$--$10^9 M_{\Sun}$ ranges.
\end{abstract}

\begin{keyword}
cosmology: dark matter\sep Galaxy: evolution\sep Galaxy: halo -- Galaxy:
kinematics and dynamics\sep Galaxy: structure
\end{keyword}

\end{frontmatter}

%\linenumbers

%%%%%%%%%%%%%%%%%%%%%%%%%%%%%%%%%%%%%%%%%%%%%%%%%%

%%%%%%%%%%%%%%%%% BODY OF PAPER %%%%%%%%%%%%%%%%%%

\section{Introduction}

The nature of dark matter (DM) remains one of the most elusive mysteries of
modern cosmology. While the existence of dark matter can be inferred from
multiple observations, from the CMB to dwarf galaxies, its mass and
interactions, even its corpuscular-versus-wave nature, is still unknown. In
the last few years a new contender, primordial black holes (PBH) dark matter,
see e.g.~\cite{Garcia-Bellido:2017fdg}, has been reborn thanks to the LIGO
detection of GW from massive BBH coalescence~\citep{Abbott:2016nmj}. Their
origin from the collapse of primordial fluctuations is rather
generic~\citep{GarciaBellido:1996qt}, but their mass range and clustering
properties are mostly uncertain. Some scenarios predict large non-Gaussian
tails in the matter density contrast distribution that could give rise to
clusters of PBH~\citep{Ezquiaga:2019ftu}, thus evading all of the
observational constraints~\citep{Clesse:2017bsw,Carr:2019kxo}. The abundance,
size and mass of these clusters is rather model dependent. What we can infer
from N-body simulations is their density profile and their spatial
distribution over the halos of galaxies~\citep{Trashorras:2020mwn}. Their
behaviour on large scales is indistinguishable from that of the usual
collisionless dark matter components of cosmological N-body simulations that
agree remarkably well with cosmological observations \citep{Zavala:2019gpq}.
What differs from the usual particle dark matter (PDM) scenario is the small
scale structure, and particularly the abundance and properties of subhalos
with masses below $10^6~M_\odot$. Many of these PBH clusters evaporate via
slingshot effects over the age of the universe, leaving dense cores of
$10^3~M_\odot$~\citep{Trashorras:2020mwn}. These small-mass PBH clusters orbit
around the galaxy and interact with other collapsed objects like globular
clusters and dwarf galaxies (see \cite{Montanari:2019rvu} for a prospective
analysis constraining such interactions via trajectory intersections with
Milky Way hyper-velocities stars). When these are tidally stretched by
successive passages through the disk of the galaxy, they become what is known
as tidal streams, almost one-dimensional concentrations of stars that stretch
across the
sky~\citep{1988ApJ...325..583A,1994MNRAS.270..209M,1995ApJ...451..598J}. These
highly-extended stellar streams are excellent targets for dark compact
clusters of stars or black holes orbiting the central halo of the
galaxy~\citep{Ibata:2001iv,Banik:2019cza,Bonaca:2018fek}. In this work we
explore the possibility to distinguish and classify by mass the different
perturbers that populate the galactic halo. In the future we may add to this
analysis their size and concentration. Our ability to distinguish low-mass
from high-mass clusters will then allow us to obtain information about the
nature of dark matter. For instance, both Fuzzy DM and Warm DM predict that
the smallest compact structures should have masses of order $10^7~M_\odot$,
while axion miniclusters are typically in the range of $10^{-9}~M_\odot$.
While PDM is expected to cluster down to
$10^{-6}~M_\odot$~\citep{Bertschinger:2006nq}, we don't expect a peak
abundance of DM clusters in the $10^3-10^5~M_\odot$ range. Therefore, a
detection of density perturbations in stellar streams suggesting an excess of
subhalos compared to PDM would be an indication of the PBH nature of DM. On
the other hand, these intermediate mass scales are too small to be detected by
strong gravitational lensing effects on the light of distant galaxies and too
large to significantly affect the microlensing of nearby stars. Therefore, we
propose these methods as a new way to explore the nature of DM.

A convolutional neural network was built by \cite{Petac:2019lam} to classify
DM subhalo masses based on Milky Way stars phase-space dynamics, expecting to
be able to constrain masses down to $10^7 M_{\Sun}$ in the near future. This
lower limit is set in large part by the complicated stellar physics around the
disk. In this work we focus on the Milky Way halo, and study stellar tidal
stream dynamics in a regime where perturbation theory in the action-angle
representation of streams applies \citep{Bovy:2016irg}. The simplified picture
allows us to reach high classification accuracy even for lower masses.

The density power spectrum of tidal streams was then studied
by~\cite{Bovy:2016irg,Banik:2018pjp,Banik:2018pal,Banik:2019smi,Banik:2019cza,Dalal:2020mjw},
constraining the properties of perturbers for the Palomar 5 and GD-1 streams.
\cite{Bovy:2016irg} showed that not only the density, but also the mean track
power spectrum and their bispectra contain important complementary
information. The bispectrum is sensitive to profile asymmetries that are
missed by the power spectrum. Given the non-Gaussian nature of density
profiles, higher order correlation functions are likely to contain relevant
information. Furthermore, given the computational requirements needed to
estimate the Bayesian evidence based on the power spectrum, literature results
relied on approximate Bayesian computations. \cite{Hermans:2020skz} proposed a
likelihood-free Bayesian inference pipeline built upon neural networks that
avoids possibly insufficient summary statistic. In this work we also aim at
profiting from the fully non-Gaussian information of stellar streams density
profiles using machine learning techniques, but also a much simpler
Bayesian approach.

\cite{Bonaca:2018fek,Bonaca:2020psc} showed that the gaps and the spur
observed in GD-1 are consistent with having originated from a collision with a
single massive perturber, see also~\cite{Gialluca:2020tno} and
~\cite{2021MNRAS.501..179M}. In this work we use a simulated GD-1-like tidal
stream allowing the possibility of several impacts. Indeed, multiple impacts
are expected in the standard cold dark matter (CDM) scenario, which for a
GD-1-like stream amount to $\sim 60$ for perturbers in the range of masses
$10^5$--$10^9 M_{\Sun}$ \citep{2016MNRAS.463..102E}. This approach has the
advantage of profiting from information coming not only from manifest features
like large gaps, but from the full star distribution within the stream.

We present two novel methodologies aimed at constraining the mass of stellar
streams dark matter perturbers. First, we outline a Bayesian model selection
pipeline based on the PDF of stellar streams density perturbations. The PDF is
a straightforward observable that takes into account the fully non-Gaussian
information of density perturbations. For instance, it has been proven to be a
valuable tool to model non-Gaussian statistics in Large Scale Structure
studies
\citep{Patton:2016umg,Gruen:2017xjj,Salvador:2018kvx,Uhlemann:2019gni,Zurcher:2020dvu},
competitive but less computationally requiring than standard analyses based on
correlation functions. Second, we consider a very different and complementary
approach by training a gradient boosting classifier. As an application, we run
simulations consistent with GD-1-like tidal stream and we forecast the
possibility of selecting which one between different perturber mass ranges is
favored by data. We consider a small $10^3$--$10^5 M_{\Sun}$, and a large
$10^5$--$10^9 M_{\Sun}$ mass range. The smallest mass range is so far
unexplored and motivated by the fact that it is suitable to test the PBH CDM
scenario which is more likely to form subhalo objects of such masses than
vanilla CDM \citep{Garcia-Bellido:2017fdg}. Given uncertainties in the
theoretical modeling of PBH, as a further test for the discriminating power of
the analyses we also discuss results obtained splitting the largest mass range
into two $10^5$--$10^7 M_{\Sun}$ and $10^7$--$10^9 M_{\Sun}$ ranges. This
analysis will motivate future studies based on the specific statistical
differences between PBH CDM and other DM models over the full $10^3$--$10^9
M_{\Sun}$ mass range.

In \autoref{sec:sims} we describe our simulated stellar streams. In
\autoref{sec:methods} we outline the two methodologies proposed to constrain
DM clusters masses and show results in \autoref{sec:results}. We conclude in
\autoref{sec:conclusions}. \ref{sec:convergence} discusses convergence
properties for our simulations. \ref{sec:powspec} considers the power spectrum
of the streams and its limitations for a model selection analysis.

\section{Simulations}
\label{sec:sims}

We model a smooth GD-1-like tidal stream relying on the line-of-parallel-angle
approach developed in \cite{Bovy:2016irg}, based on the action-angle
representation of streams \citep{Bovy:2014yba,2016MNRAS.457.3817S}. In this
formalism statistical properties of cold streams are well described by the
density $\rho$ and mean track of parallel frequency $\langle
\Omega_{\parallel} \rangle$ (average frequency along the stream) as a function
the angle offset $\Delta\theta_{\parallel}$ (parallel to the direction along
which the stream stretches) from the progenitor. The transformation from
angle-frequency space in the neighbourhood of the stream to observable
configuration space can be computed efficiently \citep{Bovy:2014yba}, allowing
to assess the effect of observational errors.

Numerical simulations rely on
\texttt{galpy}\footnote{\url{https://github.com/jobovy/galpy}.}
\citep{Bovy:2014vfa} and its extension
\texttt{streamgap-pepper}\footnote{\url{https://github.com/jobovy/streamgap-pepper}.}
\citep{Bovy:2016irg}. Statistical sampling of multiple impacts follows section
2.3 of \cite{Bovy:2016irg} to which we refer for more details and that here
we only briefly review focusing on aspects that are relevant when
considering PBH CDM perturbers rather than vanilla CDM. We neglect
perturbations induced by baryonic structures as they are expected to be
subdominant for a GD-1-like stream \citep{Hermans:2020skz}. The following
quantities are sampled sequentially:
\begin{enumerate}
\item Impact time $t_i$. We sample 64 values of time, corresponding to an
  interval of $\sim 140$ Myr (smaller than the radial period $\sim 400$ Myr of
  the stream) given the 9 Gyr age of the stream.\footnote{\label{fn:len} We
    set this age to compare our results to those of \cite{Bovy:2016irg} (see
    \autoref{sec:convergence}). More recent results suggest a lower fiducial
    value of 3.4~Gyr \citep{2019MNRAS.485.5929W}. As the stream length is
    proportional to the star velocity dispersion times its age, a younger
    stream (with the same length as the older one) would imply a larger
    velocity dispersion. However, the age estimate is very uncertain, our
    reference value of 9~Gyr being an upper limit. The age should eventually
    be marginalized over to compare with observations \citep{Banik:2019cza}.}
\item Angular offset $\Delta\theta_{i \parallel}$from the progenitor
  of closest approach along the stream.
\item Fly-by velocity ${\bf w}$ of the perturber, assuming a Gaussian
  distribution of CDM subhalos with dispersion $\sigma_h$
  \citep{2016MNRAS.463..102E}).
\item \label{item:mass} Mass $M$ and internal structure of the perturber. The
  internal profile is described by the scale parameter $r_s$ of an Hernquist
  profile (see \ref{sec:convergence} for more details and a
  comparison with a Plummer profile).\footnote{The internal profile in the PBH
    CDM scenario differs from Hernquist and Plummer profiles as it shows a
    spike towards the center \citep{Trashorras:2020mwn}. However, this
    deviation only concerns a narrow innermost region of the perturber and
    does not affect our results.} The mass $M$ probability distribution is
  taken to be $p(M) \propto r_s(M)dn_h/dM$, where we further assume a CDM
  subhalo spectrum $dn_h/dM \propto M^{-2}$ \citep{2011ApJ...731...58Y} and a
  scale factor obeying the deterministic relation $r_s(M) \propto M^{0.5}$
  \citep{2008Natur.454..735D}.
\item Impact parameter $b$, uniformly sampled along the diameter of a
  cylindrical volume with radius $b_{\rm max}$ around the stream
  \citep{2016MNRAS.463..102E}. Setting the radius proportional to the scale
  parameter, $b_{\rm max}=Xr_s$, takes effectively into account that
  decreasing the perturber mass also the volume of interaction decreases. We
  set $X=5$ as a compromise between computational convenience and numerical
  convergence of results (see \ref{sec:convergence}).
\item Number of impacts since the beginning of stream disruption to today,
  sampled from a Poisson distribution given the expected number $N_i =
  \sqrt{\frac{\pi}{2}} r_{\rm avg} \sigma_h t_d^2 \Delta\Omega^m b_{\rm max}
  n_h$, where $r_{\rm avg}$ is the average spherical radius of the stream,
  $t_d$ is the time corresponding to the start of the stream disruption,
  $\Delta\Omega^m$ is the mean parallel frequency parameter of the smooth
  stream, and $n_h$ is the number of CDM sub-halos in the given mass range,
  estimated by extrapolating results compatible with the Via Lactea II
  simulation in the $10^6$--$10^7$ mass range
  \citep{2011ApJ...731...58Y,Bovy:2016irg}. Given the reference impact
  parameter factor $X=5$, this brings $n_h \sim 839$ and $n_h \sim 62$ in the
  $10^3$--$10^5 M_{\Sun}$ and $10^5$--$10^9 M_{\Sun}$ mass ranges of our
  interest, respectively.
\end{enumerate}
Note that the effect of the dark matter encounters are well modeled in the
impulse approximation as instantaneous velocity kicks \citep{Bovy:2016irg},
neglecting effects of the collision on the perturber properties. To summarize,
we sample from the following probability distribution:
\begin{equation}
  \label{eq:p_samp}
  \begin{split}
    p(t_i, \Delta\theta_{i \parallel}, {\bf w}, M, r_s, b, N_i) = & \
    p(t_i) \
    p(\Delta\theta_{i \parallel} | t_i) \
    p({\bf w} | t_i, \Delta\theta_{i \parallel}) \\
    & p(M, r_s | t_i, \Delta\theta_{i \parallel}, {\bf w}) \\
    & p(b | t_i, \Delta\theta_{i \parallel}, {\bf w}, M, r_s) \\
    & p(N_i | t_i, \Delta\theta_{i \parallel}, {\bf w}, M, r_s, b)
  \end{split}
\end{equation}
For each sample we retrieve the density profile as a function of the parallel
angle offset from the progenitor $\Delta\theta_{\parallel}$.

\begin{figure}
  \centering
  \includegraphics[width=0.49\textwidth]{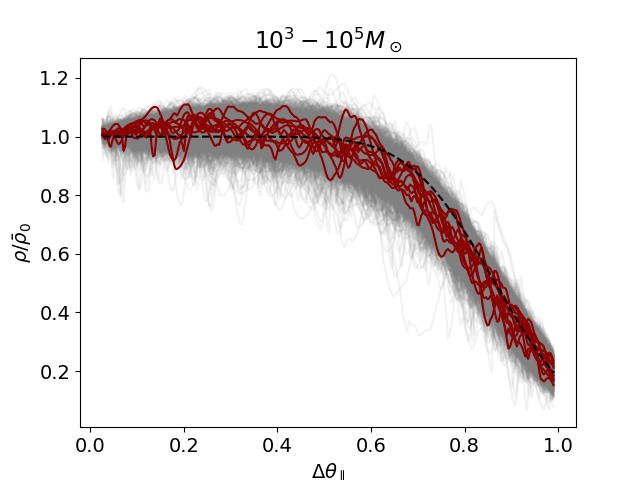}
  \includegraphics[width=0.49\textwidth]{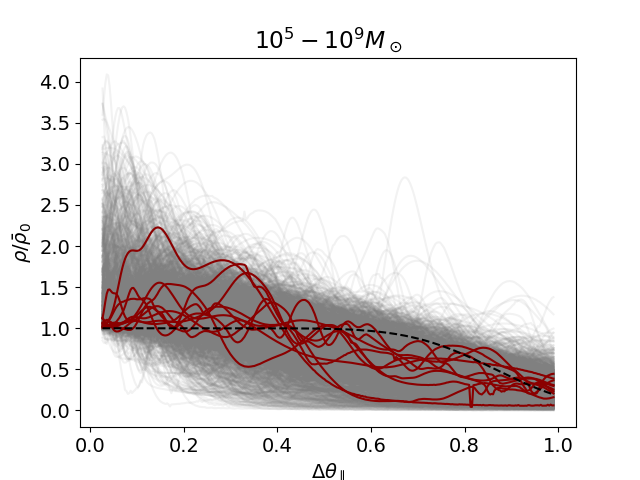}
  \caption{Simulated density profiles as a function of the parallel
    angle offset from the progenitor for different perturbers mass
    ranges. Dashed lines are the smooth density profiles. Grey lines
    are the perturbed density profiles. Red lines are a random
    selection of the latter with the purpose of visually distinguish a
    few full profiles.}
  \label{fig:rho}
\end{figure}

\begin{figure}
  \centering
  \includegraphics[width=0.49\textwidth]{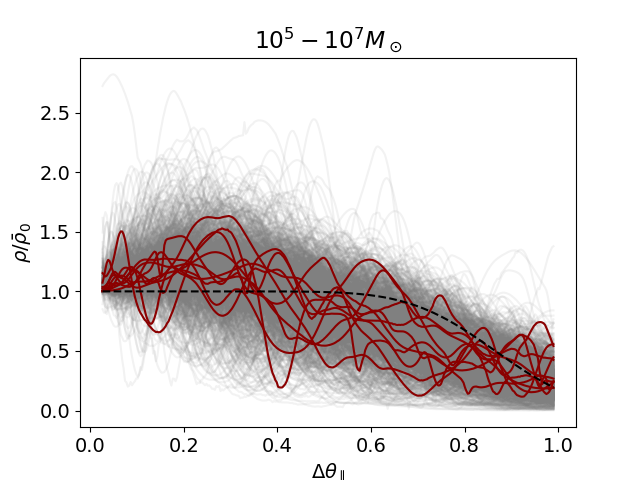}
  \includegraphics[width=0.49\textwidth]{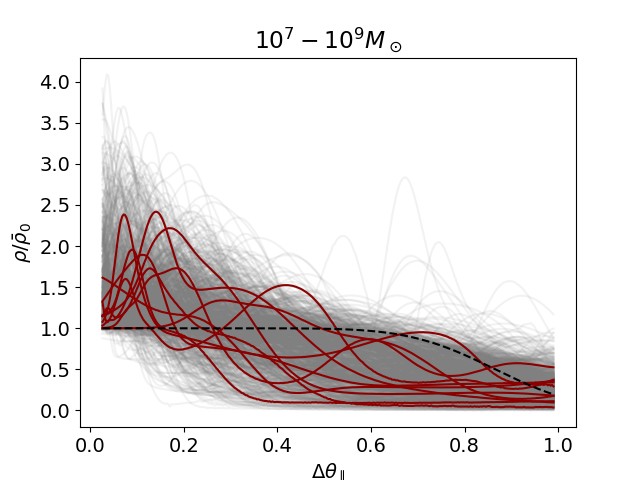}
  \caption{Like \autoref{fig:rho}, but separating the large mass range
    $10^5$--$10^9 M_{\Sun}$ into two classes.}
  \label{fig:rho_largemass}
\end{figure}

We compute at least 1000 samples of perturbed density profiles for each mass
range $10^3$--$10^5 M_{\Sun}$, $10^5$--$10^9 M_{\Sun}$, shown in
\autoref{fig:rho} normalized such that the smooth profile is $\bar\rho_0 = 1$
close to the progenitor (see e.g. Figure 1 of \cite{Bovy:2016irg} for more
details about the unperturbed stream as a function of observable Galactic
longitude). The smooth density profile is roughly constant up to
$\Delta\theta_\parallel \approx 0.6$ and then it decreases. As expected,
smaller mass perturbers lead to smaller deviations from the smooth profile
$\bar\rho$. More precisely, perturbations associated to $10^3$--$10^5
M_{\Sun}$ perturbers are at the $\sim 10\%$ level (larger only towards the
tail), while those induced by $10^5$--$10^9 M_{\Sun}$ perturbers are much
larger, rescaling $\rho$ up to a factor $\sim 4$. The perturbations profile is
peaked towards the progenitor location $\Delta\theta_\parallel \to 0$ for
large mass perturbers (typically leading to $\rho/\bar\rho > 1$), and strongly
skewed towards $\rho/\bar\rho < 1$ further away from the progenitor;
\autoref{fig:rho_largemass} shows that the peak towards
$\Delta\theta_\parallel \to 0$ is due to pertubers of masses larger than $10^7
M_{\Sun}$. Instead, perturbations induced by smaller mass objects are not
peaked towards $\Delta\theta_\parallel \to 0$ and they are distributed more
symmetrically around $\bar\rho$ up to the decrease at $\Delta\theta_\parallel
\approx 0.6$. This suggests that perturbers induce a gravitational drag
towards the center of the stream, but the large number of impacts
corresponding to the lower mass ranges symmetrizes the distribution around the
smooth stream. Gravitational drag towards the center of the stream also
induces skewness towards small density values at the already low density tail
$\Delta\theta_\parallel > 0.6$ for the small mass range. As shown later in our
results, these very different perturbed profiles makes it possible to easily
classify the three perturbers mass ranges here considered.

Note that for the low mass range the $\Delta\theta_\parallel$ corresponding to
the density break of the smooth stream could be constrained within model
uncertainties. However, in our simulations we fix the length of the stream and
set its velocity dispersion based on its age (see also footnote~\ref{fn:len} and
\autoref{sec:convergence}).

The Jacobian of the change of coordinates between the parallel angle
$\Delta\Omega_{\parallel}$ and the Galactic longitude $l$ \citep{Bovy:2016irg}
allows us to transform simulated density profiles
$\rho(\Delta\Omega_{\parallel})$ to functions $\rho(l)$ of the observable
Galactic longitude $l$. Our target is the density contrast normalized to the
smooth profile, $Q\rho(l) = \rho(l) / \bar\rho(l)$,\footnote{Here we introduce
  the quotient operator $Qx=x_1/x_0$, analog to the usual difference operator
  $\Delta x = x_1-x_0$ \citep{Henrici1964}. Note that \cite{Bovy:2016irg}
  defines instead $\delta = \rho/\bar\rho$, but we avoid confusion with the
  common $\delta = \left( \rho - \bar\rho \right) / \bar\rho$ notation used in
  cosmology.} binned in longitude. \footnote{$Q\rho(l)$ can be measured from
  observations following, e.g., the methodology of \cite{Banik:2019cza} or
  \cite{Hermans:2020skz}.} The number $n_l \sim 300$ of angular bins spanning
over $\Delta l \approx 70^{\circ}$ gives a resolution of $\sim 0.2^{\circ}$.
We assume Poisson shot noise $\epsilon(l) = 1/\sqrt{n(l)}$, where $n(l)$ is
the star density per longitude. Following \cite{Bovy:2016irg} we approximate
shot noise to be 10\% of the smooth density profile (see also
\cite{Banik:2018pjp,Banik:2019cza} that show similar errors for angular bins
along the stream), contributing to uncertainties in $Q\rho$ as
$\sigma_{\epsilon}=0.1$. Such an error is included by re-sampling density
profiles from normal distributions with mean $\rho(l)$ and standard deviation
$\sigma_{\epsilon}$ for each $l$.\footnote{Note that \emph{sampled} values can
  be negative $\rho(l)<0$, although the \emph{expectation} value for the
  density must be positive by definition.} While the shot noise level is
comparable to the size of perturbations for the lower mass range shown in
\autoref{fig:rho} (which does not include shot noise), it does not prevent us
from discerning perturbations induced by large mass perturbers and hence to
perform model selection. Note that such a level of shot noise alleviates
errors due to extrapolation of N-body results down to the so far unexplored
$10^3$--$10^5 M_{\Sun}$ clusters range.

\section{Methods}
\label{sec:methods}

In this section we outline two different methodologies for model selection: a
simple Bayesian pipeline based on the PDF, and a gradient boosting classifier.

\subsection{Bayesian model selection based on PDF}
\label{sec:pdf}
Here we outline a methodology to perform model selection based on Bayesian
principles \citep{Trotta:2017wnx}. \cite{Bovy:2016irg} computed the power
spectrum and the bispectrum of the density profile, showing that both probes
contain valuable complementary information. Indeed, density perturbations are
not expected to follow Gaussian profiles and even higher-order correlations than
the bispectrum may contain valuable information. Therefore, rather than
considering polyspectra, we compute the PDF of the density contrast profile
$Q\rho(l)$.

\subsubsection{Modeling the prior distribution}
\label{sec:prior}

\begin{figure}
  \centering
  \includegraphics[width=0.49\textwidth]{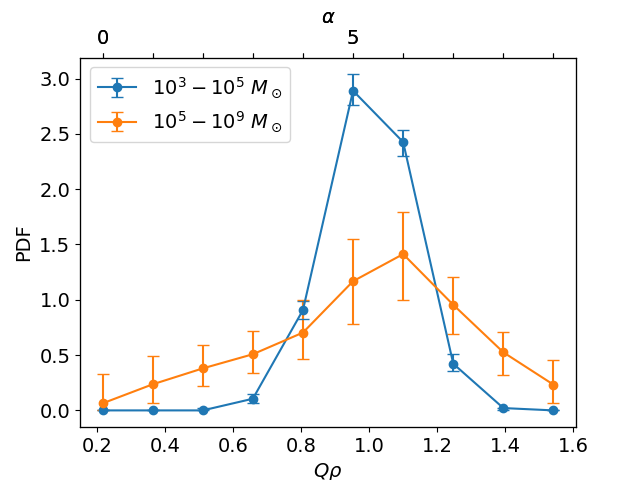}
  \includegraphics[width=0.49\textwidth]{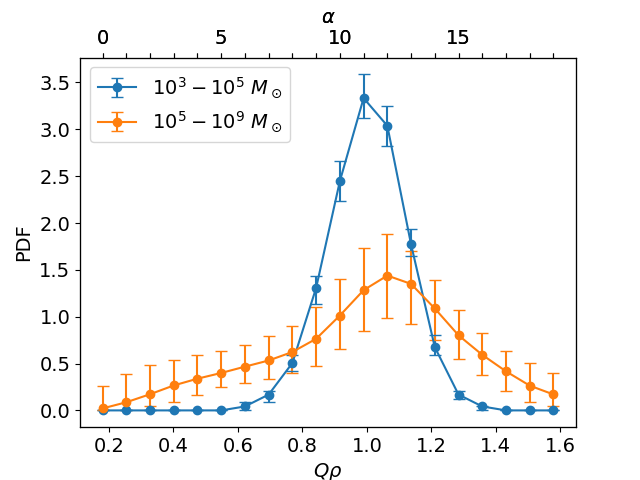}
  \includegraphics[width=0.49\textwidth]{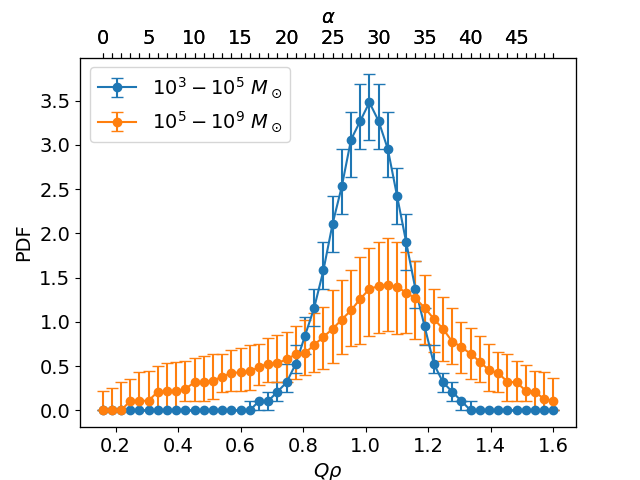}
  \includegraphics[width=0.49\textwidth]{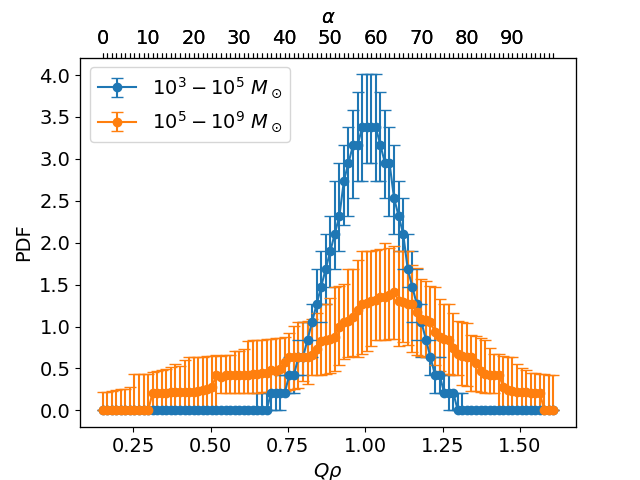}
  \caption{Interquantile ranges and medians for PDF bins of $Q\rho =
    \rho/\bar\rho$ profiles measured from simulations. Panels correspond to
    $N_{\rm bin} = 10, 20, 50, 100$ (top axes shows the bin number).}
  \label{fig:pdferr}
\end{figure}

We compute histograms of simulated density profiles $Q\rho_i(l) =
\rho_i(l)/\bar\rho_i(l)$ obtained sampling from \autoref{eq:p_samp}. Given the
counts $c_{\alpha}$ for the $\alpha^{th}$ histogram bin of width $w_{\alpha}$,
the respective density is obtained as $y_{\alpha} =
\frac{c_{\alpha}}{w_{\alpha} \sum_{\beta} c_{\beta}}$, where the normalization
is such that the integral of the distribution is equal to one.\footnote{To
  avoid confusion with longitude bins $l_j$ and with references to a given
  $i^{th}$ sample of our simulations, we denote PDF histogram bins with Greek
  indices in the range $\alpha = 1, \ldots, N_{\rm bin}$.} This determines the
probability density function ${\rm PDF}(Q\rho)$. \autoref{fig:pdferr} shows
median values and interquantile ranges obtained computing the ${\rm
  PDF}(Q\rho)$ for all simulations and for different choices of the number of
bins $N_{\rm bin}$. The range in $Q\rho$ is set to be the largest one where we
have samples for both mass ranges. The ${\rm PDF}(Q\rho)$ functions are peaked
around $Q\rho \approx 1$, but consistently with \autoref{fig:rho} the larger
mass ranges have broader profiles. The ${\rm PDF}(Q\rho)$ for the lower mass
range is close to a normal distribution with mean $Q\rho=1$ and standard
deviation $\sigma=\sigma_{\epsilon}=0.1$, consistently with the fact that it
is strongly affected by shot noise $\epsilon$. Neglecting shot noise would
enhance negative skewness.

\begin{figure*}
  \centering
  \includegraphics[width=0.24\textwidth]{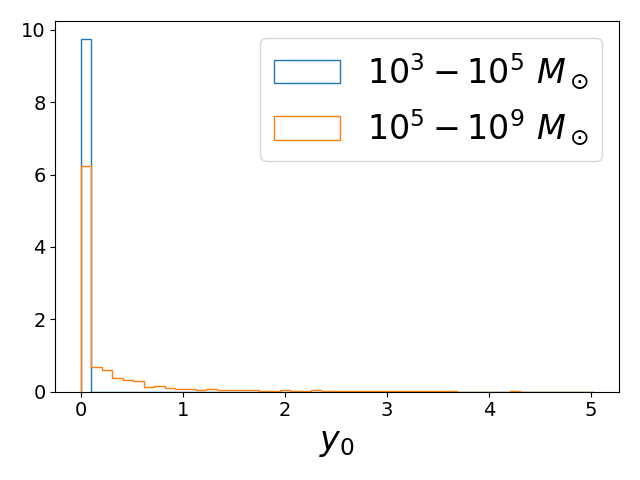}
  \includegraphics[width=0.24\textwidth]{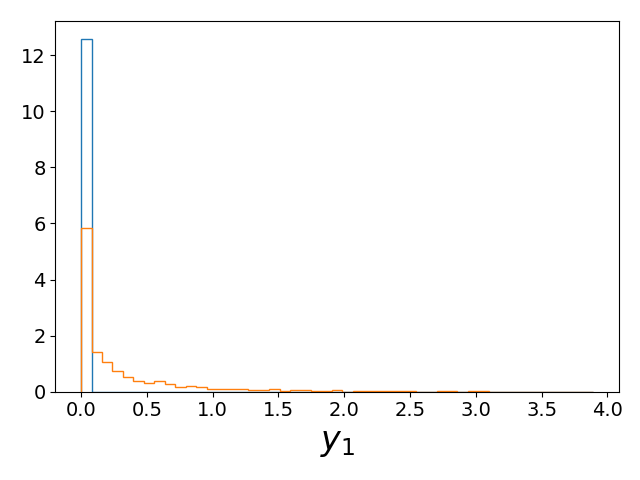}
  \includegraphics[width=0.24\textwidth]{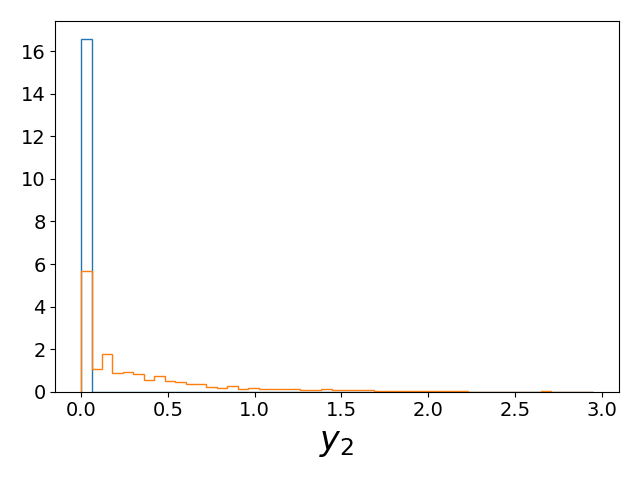}
  \includegraphics[width=0.24\textwidth]{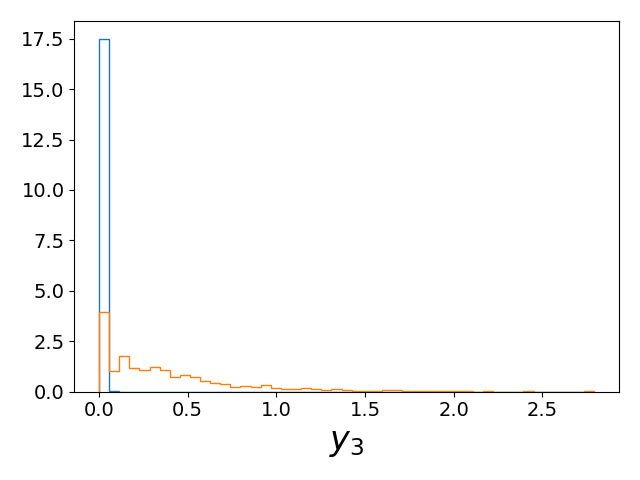}
  \\
  \includegraphics[width=0.24\textwidth]{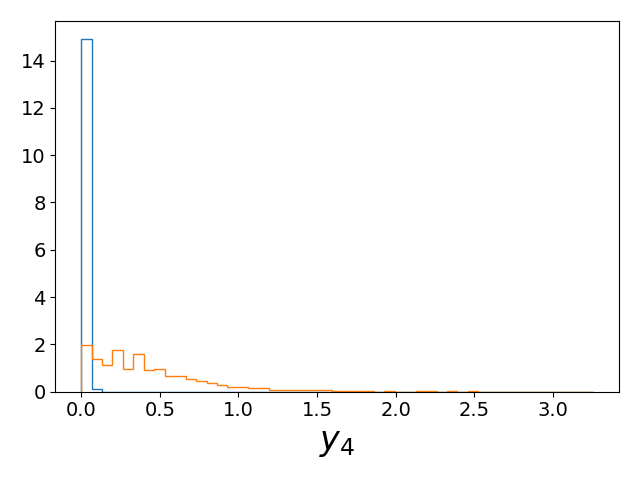}
  \includegraphics[width=0.24\textwidth]{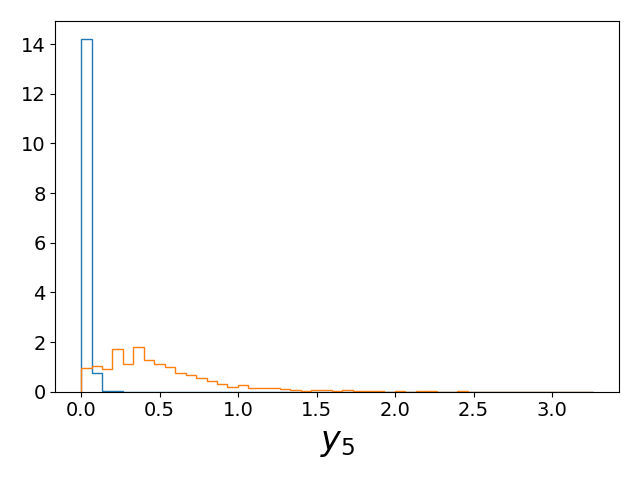}
  \includegraphics[width=0.24\textwidth]{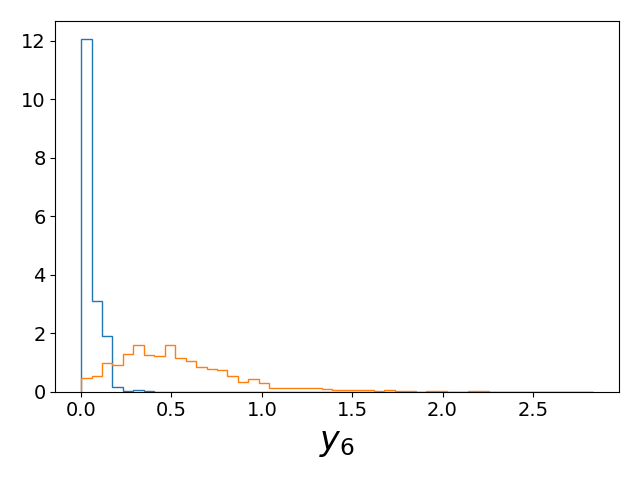}
  \includegraphics[width=0.24\textwidth]{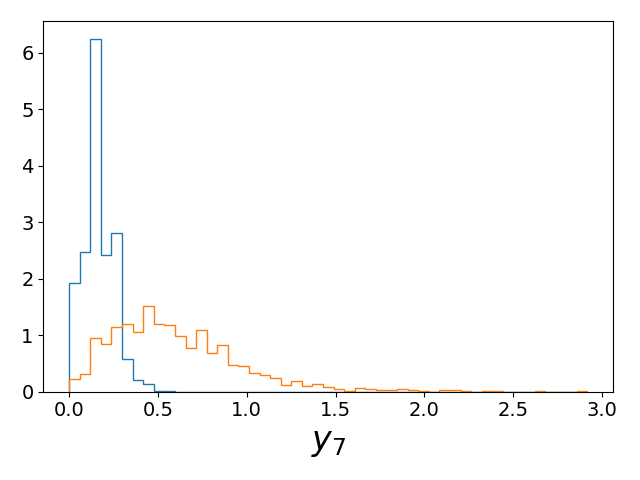}
  \\
  \includegraphics[width=0.24\textwidth]{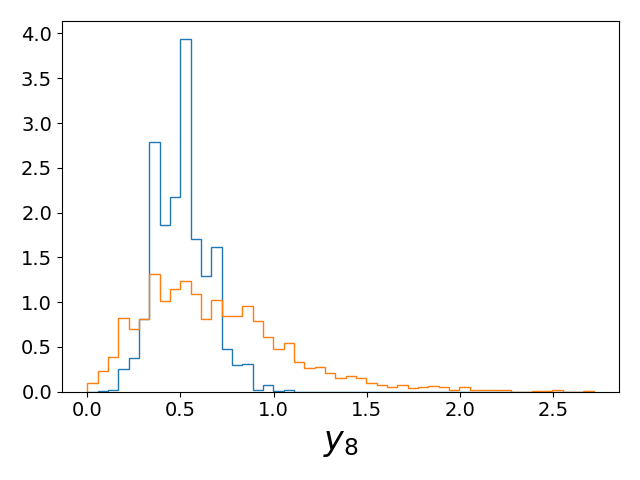}
  \includegraphics[width=0.24\textwidth]{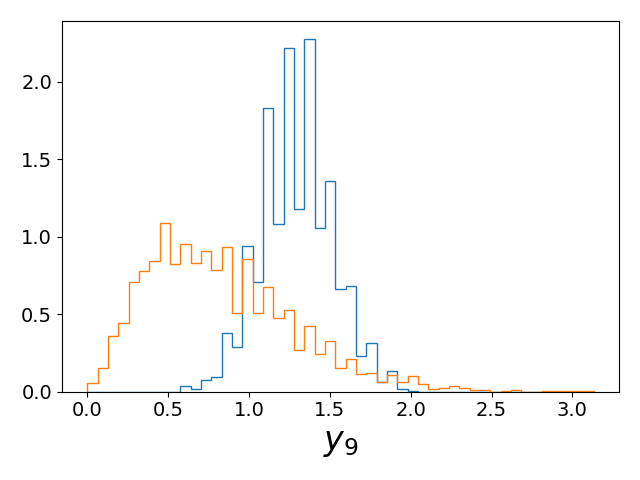}
  \includegraphics[width=0.24\textwidth]{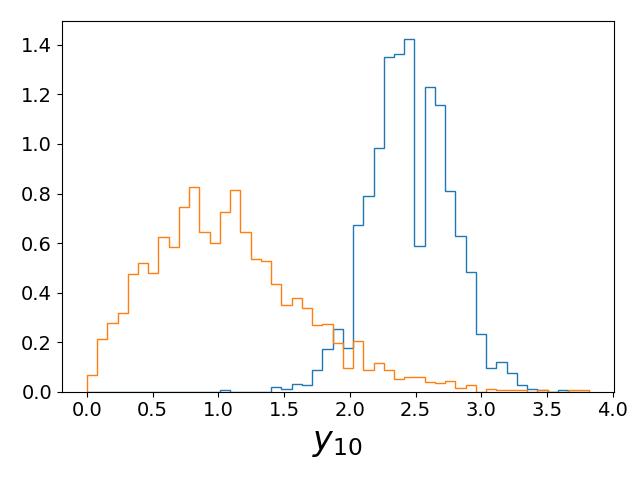}
  \includegraphics[width=0.24\textwidth]{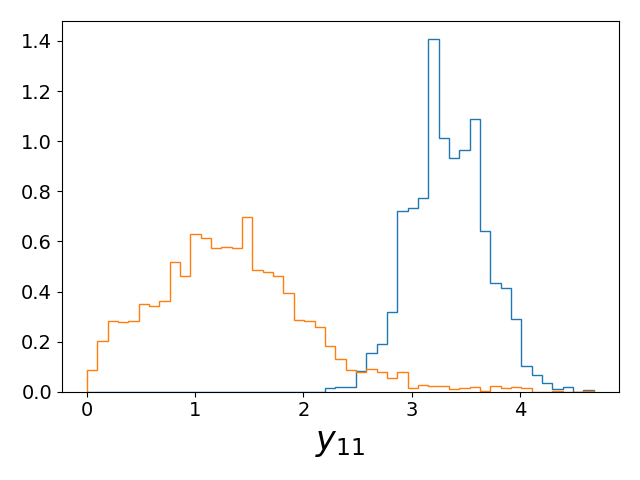}
  \\
  \includegraphics[width=0.24\textwidth]{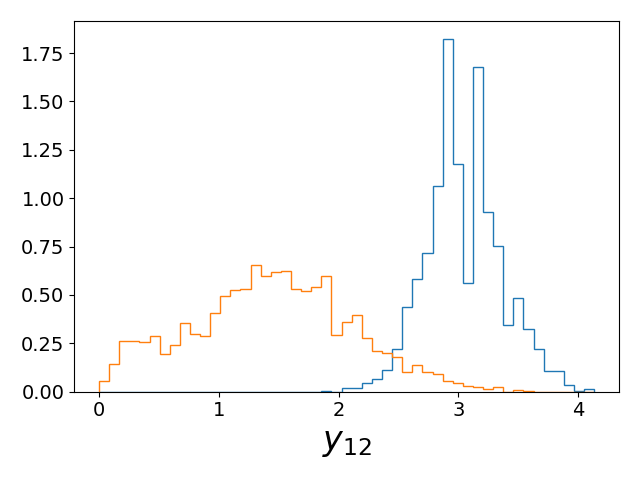}
  \includegraphics[width=0.24\textwidth]{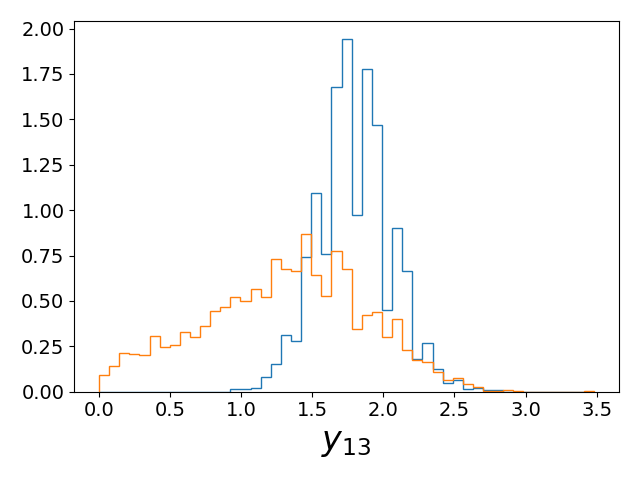}
  \includegraphics[width=0.24\textwidth]{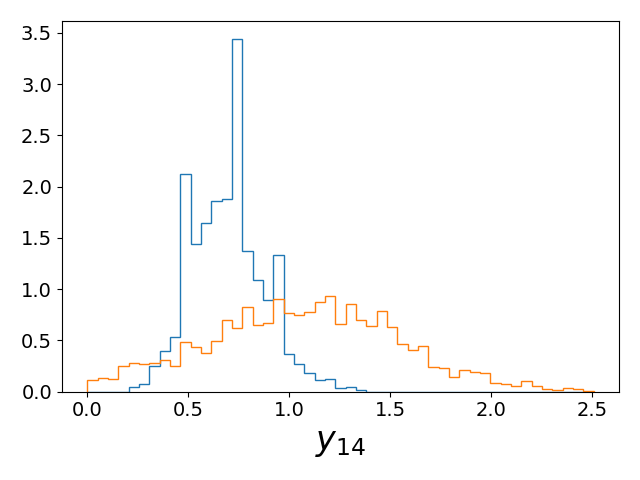}
  \includegraphics[width=0.24\textwidth]{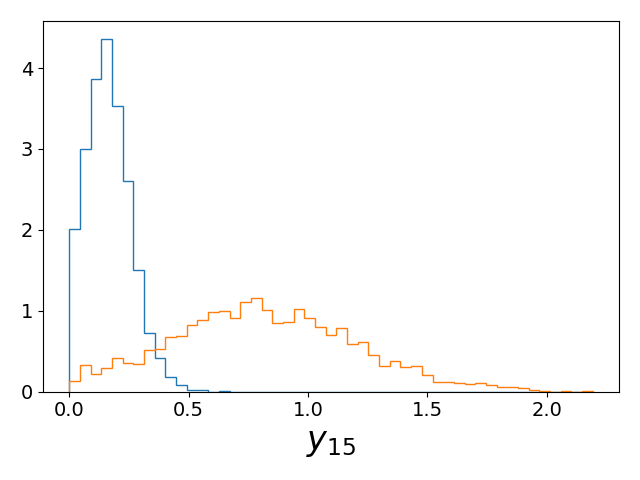}
  \\
  \includegraphics[width=0.24\textwidth]{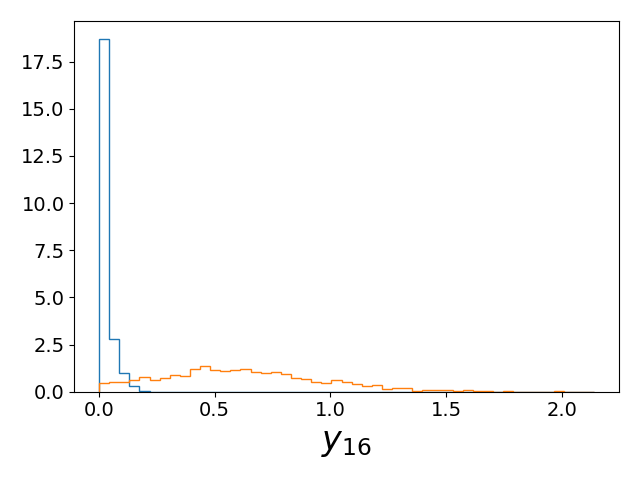}
  \includegraphics[width=0.24\textwidth]{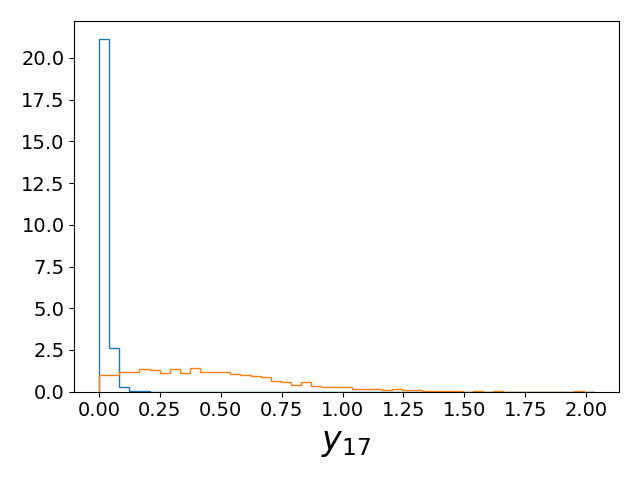}
  \includegraphics[width=0.24\textwidth]{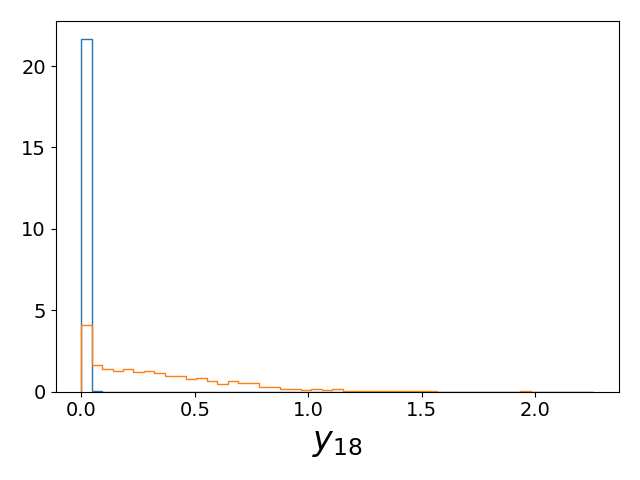}
  \includegraphics[width=0.24\textwidth]{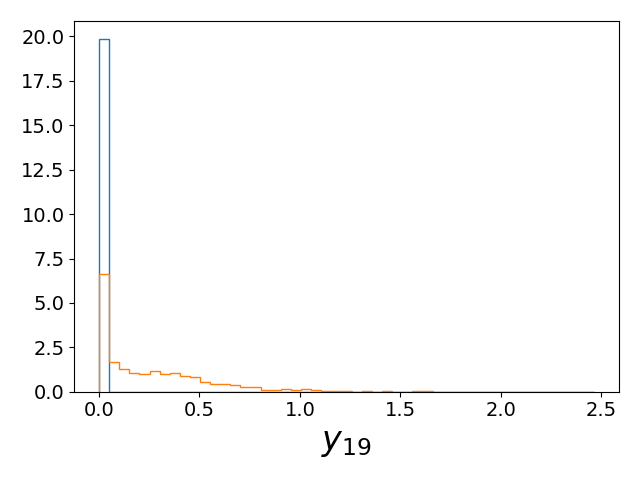}
  \caption{Panels show $y_{\alpha}$ distributions, namely the PDFs of each ${\rm
      PDF}(Q\rho)$ bin $\alpha$ for the $N_{\rm bin}=20$ case (see
    \autoref{fig:pdferr} for comparison). Here we recover each $y_{\alpha}$
    using 50 bins, but when performing model selection we marginalize over
    10--500 bins. In panels corresponding to the tails of ${\rm PDF}(Q\rho)$
    (i.e., small and large $\alpha$ values) we do not have enough resolution to
    estimate the PDF of the small mass case.}
  \label{fig:pdf_pdf}
\end{figure*}

The PDF of each ${\rm PDF}(Q\rho)$ bin $y_{\alpha}$---i.e., ${\rm
  PDF}(y_{\alpha})$---provides an estimate of the probability distribution
$\Pi( y_{\alpha} | \mathcal{M})$ for model $\mathcal{M}$. Our models
$\mathcal{M}_0$ and $\mathcal{M}_1$ correspond to the $10^3$--$10^5 M_{\Sun}$
and $10^5$--$10^9 M_{\Sun}$ mass ranges, respectively. The $\Pi(y_{\alpha} |
\mathcal{M})$ distributions so recovered based on simulations serves as prior
for model selection. \autoref{fig:pdf_pdf} shows ${\rm PDF}(y_{\alpha})$ for
all models. The small mass perturbers model is characterized by tiny
$y_{\alpha}$ values for $\alpha$ bins corresponding to the tails of ${\rm
  PDF}(Q\rho)$ (see \autoref{fig:pdferr}), while all mass models show a
log-normal distribution for $\alpha$ bins corresponding central regions of
${\rm PDF}(Q\rho)$.

\subsubsection{Model selection}

Let $p \left( \hat y_{\alpha} | y_{\alpha}, \mathcal{M} \right)$ be the
likelihood function for ${\rm PDF}(Q\rho)$ bin $\alpha$, where the hat denotes
data. The Bayesian evidence for a given model $\mathcal{M}$ is given by:
\begin{equation}
  \label{eq:marg-like}
  E(\hat y_{\alpha} | \mathcal{M}) = \int\ p(\hat y_{\alpha} | y_{\alpha}, \mathcal{M})
  \Pi(y_{\alpha} | \mathcal{M}) dy_{\alpha} \;,
\end{equation}
where the prior $\Pi(y_{\alpha} | \mathcal{M})$ is computed based on simulations
as outlined above. Then, model selection can rely on the Bayes factor:
\begin{equation}
  \label{eq:bayesK}
  B_{\alpha, ij} = \frac{E(\hat y_{\alpha} | \mathcal{M}_i)}
  {E(\hat y_{\alpha} | \mathcal{M}_j)} \;.
\end{equation}

Since an analysis based on true data is beyond the scope of this work, here we
rely again on simulations to estimate $p(\hat y_{\alpha} | y_{\alpha},
\mathcal{M})$ in \autoref{eq:marg-like} to forecast expected evidence values
that will motivate further investigation based on true data. We consider
different fiducial models:
\begin{equation}
  p(\hat y_{\alpha}^{{\rm fid}, i} | y_{\alpha}) = \Pi(y_{\alpha} | \mathcal{M}_i) \;,
\end{equation}
for $i=0, 1$. Note that we removed the dependence on the model on the
left-hand side, as for each fiducial case $i$ we assume the likelihood to be
the same; it is easy to relax this assumption if needed. Then we consider
\begin{equation}
  \label{eq:Eforecast}
  E(\hat y_{\alpha}^{{\rm fid}, i} | \mathcal{M}_j) =
  \int_0^{\infty} p(\hat y_{\alpha}^{{\rm fid}, i} | y_{\alpha})
  \Pi(y_{\alpha} | \mathcal{M}_j) dy_{\alpha} \;,
\end{equation}
where again $j=0, 1, 2$. This determines
\begin{equation}
  \label{eq:Bfid}
  B_{\alpha, ij}^{{\rm fid}, i} = E(\hat
  y_{\alpha}^{{\rm fid}, i} | \mathcal{M}_i) / E(\hat y_{\alpha}^{{\rm fid}, i} |
  \mathcal{M}_j) \;,
\end{equation}
which provides the strength of evidence for model $\mathcal{M}_i$ if the true
underlying mass distribution is indeed consistent with $\mathcal{M}_i$.

We sample $\Pi(y_{\alpha} | \mathcal{M}_i)$ using the same binning in
$y_{\alpha}$ for all models $\mathcal{M}_i$ so that the product of the two
distributions can be estimated by the product of the respective bin heights,
and integrals are estimated using the trapezoidal rule. Methodological errors
are estimated by marginalizing over the number of bins in $y_{\alpha}$.

\subsection{Gradient Boosting classifier}
\label{sec:gb}

We use the Python libraries \texttt{Scikit-learn} \citep{scikit-learn} and
\texttt{XGBoost} \citep{Chen:2016:XST:2939672.2939785} to train a
gradient-boosted decision trees classifier. This model builds and ensemble of
weak predictors (decision trees) to provide a robust predictor, optimizing an
objective function with an iterative gradient descent algorithm. In our
single-label (exclusive classes) binary classification case the objective is a
logistic function, whose output is the probability of the $\mathcal{M}_0$
class \citep{DLbook}.

Our data is composed by a vector $\boldsymbol{X}$ whose component $i$ is a given
simulation sample $(Q\rho_i(l_1), Q\rho_i(l_2), \ldots, Q\rho_i(l_{n_l}))$,
where $Q\rho_i(l) = \rho_i(l) / \bar\rho_i(l)$, and a vector $\boldsymbol{y}$
containing binary labels associated to the class corresponding to a given
sample. We randomly split the density samples into testing (25\% of the total
samples), training and validation (75\% and 25\% of the remaining samples,
respectively) sets. The training and validation samples are used to tune the
model hyperparameters, while the testing sample is only used to evaluate the
accuracy of the final model. Then, rather than considering all of the $n_l$
angular bins for each sample, we find it convenient to perform a Principal
Component Analysis (PCA), and only keep $n_{\rm PCA} < n_l$ components. PCA is
sensitive to the variance of the features $X_i$ that varies significantly in
size. Hence, before computing the principal components we standardize the
features by removing the mean and scaling to unit variance.

We then perform a 5-fold cross validation (CV) on the training set to optimize
the maximum tree depth \texttt{max\_depth} for base learners and the number of
gradient boosted trees \texttt{n\_estimators}. We find that optimizing the
boosting learning rate does not affect substantially our results, due to its
strong correlation with \texttt{n\_estimators}. The search grid covers a few
values for each hyperparameter (the full model has to be evaluated at each CV
fold) within the \texttt{max\_depth} $\in [2, 128]$ and \texttt{n\_estimators}
$\in [10, 200]$ ranges. The validation set is then used to evaluate the model
pipeline based on the number of true positives $T_{p_i}$, true negatives
$T_{n_i}$, false positives $F_{p_i}$ and false negatives $F_{n_i}$ for each
class $\mathcal{M}_i$. As summary metrics we consider the accuracy
\begin{equation}
  \label{eq:acc}
  A_i = \frac{T_{p_i} + T_{n_i}}{T_{p_i} + T_{n_i} +
    F_{p_i} + F_{n_i}} \;,
\end{equation}
precision
\begin{equation}
  \label{eq:precision}
  P_i = \frac{T_{p_i}}{T_{p_i} + F_{p_i}} \;,
\end{equation}
recall
\begin{equation}
  \label{eq:recall}
  R_i = \frac{T_{p_i}}{T_{p_i} + F_{n_i}} \;,
\end{equation}
and their harmonic average, or $F_1$-score
\begin{equation}
  \label{eq:F1}
  F_{1,i} = \frac{2}{1/P_i + 1/R_i} \;.
\end{equation}

We find that setting $n_{\rm PCA} = 8$ explains 80\% of the total variance and
it is a reasonable compromise between good model performance and training
speed. Setting a much lower or larger $n_{\rm PCA}$ (or not performing PCA)
would worsen model performance by a few percentage points. While we expect
that including more PCA components (or not performing PCA) should lead to at
least comparable performance, this would require to run the CV search on a
finer grid of hyperparameters. Furthermore, including more PCA components is
up to 5 times more requiring in training speed.

\section{Results}
\label{sec:results}

In this section we discuss results for both mass model selection methodologies
outlined above.

\subsection{Bayesian model selection based on PDF}
\label{sec:res_pdf}

\begin{figure}
  \centering
  \includegraphics[width=0.49\textwidth]{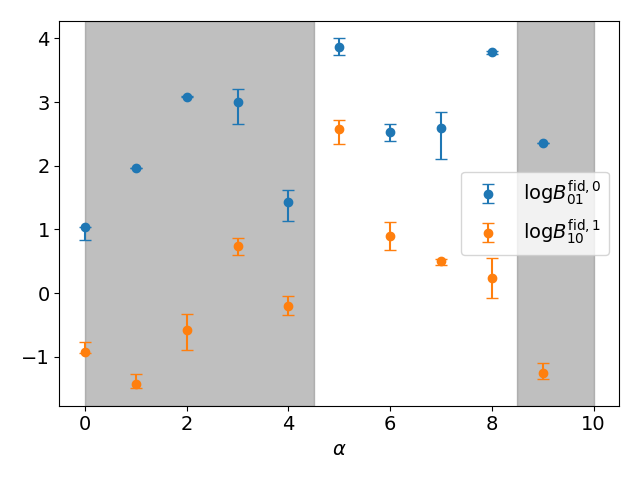}
  \includegraphics[width=0.49\textwidth]{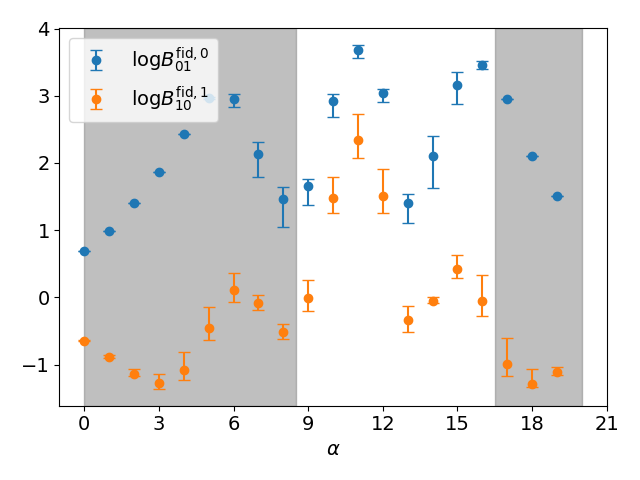}
  \includegraphics[width=0.49\textwidth]{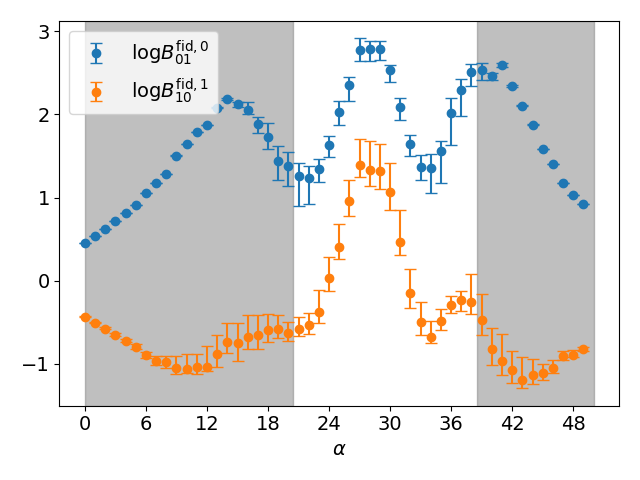}
  \includegraphics[width=0.49\textwidth]{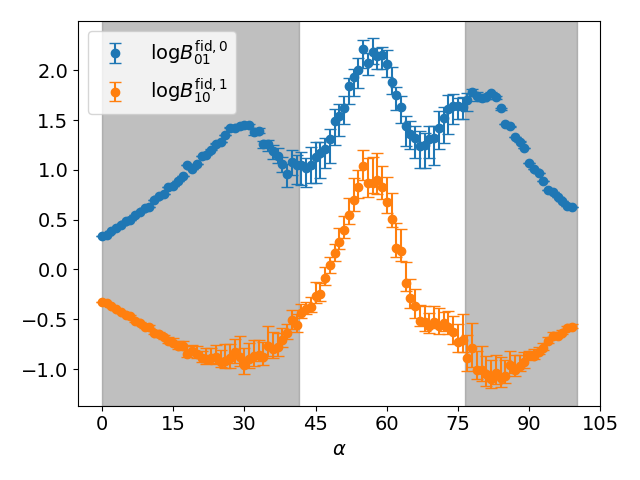}
  \caption{Bayes factors for different bins $\alpha$ of the PDF (see
    \autoref{fig:pdferr}) corresponding to $N_{\rm bin}=10, 20, 50, 100$).
    Grey regions correspond to rare density fluctuations that are poorly
    sampled by our simulations, leading to unreliable evidence estimates.}
  \label{fig:lnB}
\end{figure}

\autoref{fig:lnB} shows evidence ratios defined in \autoref{eq:Bfid} for
different choices of the number $N_{\rm bin}$ of bins in $Q\rho$. For each
${\rm PDF}(Q\rho)$ bin $y_{\alpha}$ we compute the evidence ratio selecting
different number of bins 10--500. Error bars correspond to interquantile
ranges so obtained and their narrowness confirms that our evidence estimates
based on the trapezoidal rule are robust against the binning in $y_{\alpha}$.

Results shown in \autoref{fig:lnB} are easily interpreted comparing with
\autoref{fig:pdferr} and \autoref{fig:pdf_pdf}. Central $\alpha$ bins show the
largest Bayes factor, as ${\rm PDF}(Q\rho)$ for the different models are
significantly distinct. The crossing of ${\rm PDF}(Q\rho)$ curves in
\autoref{fig:pdferr} (see also $\alpha=8, 13$ in \autoref{fig:pdf_pdf}) leads
to a decrease in $\log B$ as the models are less distinguishable, while it
increases again towards the tails where the small mass model $\mathcal{M}_0$
has ${\rm PDF}(Q\rho) \approx 0$. $\log B$ eventually decreases for the outer
bins as ${\rm PDF}(Q\rho) \to 0$ also for the larger mass model
$\mathcal{M}_1$. Bins corresponding to the tails of ${\rm PDF}(Q\rho)$
distributions do not give reliable evidence estimates due to the poor
resolution for small mass perturbers which leads to a single ${\rm
  PDF}(y_{\alpha})$ bin, see \autoref{fig:pdf_pdf}. Hereon we neglect evidence
for $\alpha$ bins outside the $\sim 3\sigma_\epsilon$ range from the peak of
the low mass ${\rm PDF}(Q\rho)$ distribution, which is the most affected one
by poor resolution.\footnote{See \autoref{sec:prior} for a discussion about
  how $\sigma_\epsilon$ affects the lower mass PDF, which justifies our choice
  here.} The excluded regions are shown as gray bands in
\autoref{fig:lnB}.\footnote{This approach has the advantage of allowing
    a robust study of the Bayes factors relying on a relative small number of
    simulations. In general, a reliable estimate of the evidence is notably
    difficult to reach due to posterior sampling requirements, see for
    instance the discussion in \autoref{sec:powspec}.}

\begin{table}
	\centering
	\caption{Cumulative Bayes factor upper limits for different choices
    of the number of ${\rm PDF}(Q\rho)$ bins.}
	\label{tab:lnB}
	\begin{tabular}{l|cccc}
		\hline
    \backslashbox{}{$N_{\rm bin}$} & 10  &  20 & 50 &  100\\
		\hline
    $\sum \log B_{01}^{{\rm fid}, 0}$ & 9.0 & 18.0 & 33.1 & 53.5 \\
    $\sum \log B_{10}^{{\rm fid}, 1}$ & 4.0  & 5.4 & 3.1 & 1.3  \\
		\hline
	\end{tabular}
\end{table}

The largest Bayes factors are in the range 2.5--4, depending on $N_{\rm bin}$,
when the fiducial model is $\mathcal{M}_0$ (see $\log B_{01}^{{\rm fid}, 0}$).
Bayes factors are smaller when the fiducial models is $\mathcal{M}_1$, as the
corresponding ${\rm PDF}(Q\rho)$ profiles (see \autoref{fig:pdferr} and
\autoref{fig:pdf_pdf}) are less sharp and characterized by larger dispersion.
While a proper estimate of cumulative Bayes factors should include their
correlation, an upper limit is given by summing $\log B_{\alpha}$ for all
$\alpha$ indices (up to the excluded region commented above) as reported in
\autoref{tab:lnB}. Assuming model $\mathcal{M}_0$ as fiducial leads to
moderate or strong evidence.\footnote{We recall that the usual Jeffreys' scale
  interprets $|\log B|$ ranges $< 1$, 1--2.5, 2.5--5 and $> 5$ as
  inconclusive, weak, moderate and strong evidence, respectively
  \cite{Trotta:2017wnx}. See, however, \cite{Nesseris:2012cq}.} Assuming model
$\mathcal{M}_1$ can instead lead to weak evidence. The largest evidences
obtained when including more bins are expected to be more strongly suppressed
by cross-bin evidence correlations that we neglect.

The analysis suggests that if CDM perturbers form clusters in the
$10^5$--$10^9 M_{\Sun}$ range it may be more difficult to exclude the
$10^3$--$10^5 M_{\Sun}$ hypothesis than the vice versa.

As an exercise to further explore the performance of the methodology, we
repeated the computation considering three mass ranges instead of two, namely
$10^3$--$10^5 M_{\Sun}$, $10^5$--$10^7 M_{\Sun}$ and $10^7$--$10^9 M_{\Sun}$.
Similarly to the case described above, we find that if CDM perturbers form
clusters in the $10^7$--$10^9 M_{\Sun}$ range it may be more difficult to
exclude the $10^3$--$10^5 M_{\Sun}$ or $10^5$--$10^7 M_{\Sun}$ hypotheses than
the vice versa, possibly leading to inconclusive results.

\subsection{Gradient Boosting classifier}

\begin{figure}
  \centering
  \includegraphics[width=0.7\textwidth]{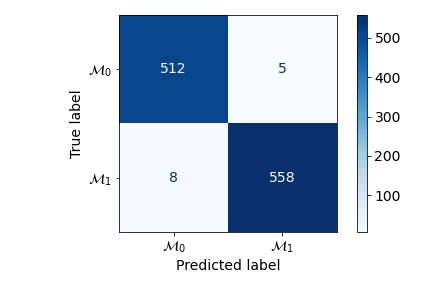}
  \caption{Confusion matrix obtained evaluating the gradient boosting model on
    the test dataset. Elements show the number of samples predicted to be
    associated to models $\mathcal{M}_0$ ($10^3$--$10^5 M_{\Sun}$ CDM
    perturbers) and $\mathcal{M}_1$ ($10^5$--$10^9 M_{\Sun}$) that are
    correctly (diagonal) or wrongly (non-diagonal) classified.}
  \label{fig:cm}
\end{figure}

\begin{table}
	\centering
	\caption{Performance, relative to each mass category, of the gradient
    boosting classifier evaluated on the test set.}
	\label{tab:gb}
	\begin{tabular}{lccc}
		\hline
    & Precision  &  Recall & $F_1$-score \\
		\hline
    $10^3$--$10^5 M_{\Sun}$   &    0.98   &   0.99  &    0.99 \\
    $10^5$--$10^9 M_{\Sun}$   &    0.99  &    0.99   &   0.99 \\
		\hline
	\end{tabular}
\end{table}

Here we use the test set described in \autoref{sec:gb} to evaluate the
performance of the model trained on the training and validation sets. Given a
$X_i$ sample from the test set, the model outputs the probabilities of it
being associated with the $\mathcal{M}_0$ class. To compute the number of
(in-)correctly classified elements we assign them to the $\mathcal{M}_0$ class
if the respective probability is larger than 50\%.

\autoref{fig:cm} shows the confusion matrix reporting the number of correct
classifications on the diagonal, and the number of wrong classifications on
non-diagonal elements. Non-diagonal elements are much smaller than the
diagonal ones. Compared to the PDF analysis, the gradient boosting algorithm
shows strong discriminating power for all masses. This is confirmed by
\autoref{tab:gb} showing the metrics described in
\autoref{eq:precision}--\autoref{eq:F1} evaluated on the test set. All classes
are characterized by $98\%$--$99\%$ scores, and also accuracy
(\autoref{eq:acc}) reaches $99\%$.

Also in this case we repeated the computation considering three mass ranges
$10^3$--$10^5 M_{\Sun}$, $10^5$--$10^7 M_{\Sun}$ and $10^7$--$10^9 M_{\Sun}$,
confirming that the metrics described in
\autoref{eq:precision}--\autoref{eq:F1} give good results $\gtrsim 90\%$ for
all classes, reaching 98\% for the smaller mass one.

\section{Conclusions}
\label{sec:conclusions}

In our search for understanding one of the most elusive mysteries of nature,
we have explored the information content associated with tidal stream
collisions with dark matter clusters in the inner halo of our galaxy. We have
introduced two complementary methodologies to constrain the mass of CDM
perturbers on stellar streams and forecasted their potential ability based on
GD-1-like stream simulations.

While the methodologies are independent from the dark matter model, we are
interested in providing a simple test to discern specifically the PBH CDM
hypothesis from the standard PDM one. If data favor stream perturbers in a
small mass $10^3$--$10^5 M_{\Sun}$ range over a larger one $10^5$--$10^9
M_{\Sun}$ (within which PDM is typically analyzed) this will be a strong hint
in favor of the PBH CDM hypothesis because a peak abundance of PBH clusters is
expected in the smaller mass range.

As no simulation is available to date for neither PBH CDM nor PDM over
$10^3$--$10^5 M_{\Sun}$ mass scales, we modeled dark matter in this regime by
extrapolating PDM results validated in the well known $10^5$--$10^9 M_{\Sun}$
regime down to smaller masses. While this is conceptually in contrast with our
simplifying hypothesis that mainly PBH CDM contributes to the $10^3$--$10^5
M_{\Sun}$ range, forecasts are still useful to characterize the envisaged
model selection capabilities and to justify further studies, especially
considering that strong shot-noise mitigates this extrapolation errors.

The Bayesian PDF model selection provides a straightforward and robust
analysis that, contrary to incomplete statistics such as the power spectrum,
takes into account the fully non-Gaussian information of stellar streams
density profiles. Including realistic observational errors, we expect large
Bayes factors when taking the low mass ($10^3$--$10^5 M_{\Sun}$, where a peak
abundance of PBH CDM is expected) perturbers as the fiducial models. Instead,
taking a large masses ($10^5$--$10^9 M_{\Sun}$) model as the fiducial one may
lead to inconclusive evidence, due to a larger dispersion in the PDF.

On the other hand, the gradient boosting model provides a complementary method
to the PDF one that allows to classify an observed density profile according
to the most likely mass range with high performance. While the PDF analysis
relies on Bayesian model selection, this is not the case for the gradient
boosting one; however, the path integral approach developed
in~\cite{Nesseris:2013bia} to compute Bayesian credible intervals for
non-parametric genetic algorithms can also be adapted to our gradient boosting
model. The excellent performance of the gradient boosting classifier also
relies on the fact that the model takes into account the ordered information
about the density profile dependence on the longitude, information that is
instead lost in the PDF analysis. \autoref{fig:rho} shows indeed an important
dependence of perturbations on the angular distance from the progenitor,
including a characteristic profile peaked towards the progenitor for large
mass perturbers (which may bring to inconclusive evidence based on the PDF),
differently from the low mass case. Another advantage of the gradient
boosting algorithm is that it is a likelihood-free approach, contrary to the
PDF analysis. While in this work we assumed different fiducial models for the
latter, applying the PDF method to observations will require a detailed
modeling of the likelihood.

To test of the robustness of the method and given theoretical uncertainties in
the PBH modeling we verified that similar conclusions are reached when
performing forecasts further dividing the largest mass range into
$10^5$--$10^7 M_{\Sun}$ and $10^7$--$10^9 M_{\Sun}$ ranges, leading to a total
of three mass ranges.

As an extension to the analysis presented here, we compared our gradient
boosting classifier with a densely connected feedforward neural network
classifier \citep{DLbook} with a few hidden layers composed by
  $\mathcal{O}(10)$ units, which gives comparable results only when
neglecting shot noise and being significantly less performing for realistic
observational errors. We also do not expected convolutional neural networks to
bring substantial improvements because, due to their translational invariance,
they would loose information about density perturbations as ordered sequences
depending on longitude. Recurrent neural networks may be an interesting
perspective that do not suffer from such a limitation, but due to typically
high computational costs and the already good performance of our gradient
boosting predictor, we don't deem it as an urgent investigation. Another
advantage of gradient boosting is that its hyperparameters are easily
optimized, avoiding empirical iterations over different neural networks
architectures.

We showed that stellar streams are a promising tool to investigate
$10^3$--$10^5 M_{\Sun}$ DM perturbers, a range outside the reach of other
observations such as lensing and of interest to discriminate the PBH CDM
hypothesis. As mentioned, our motivation of comparing this small subhalo objects mass
range with the large ones $10^5$--$10^9 M_{\Sun}$ to favor PBH CDM over
particle DM relies on the simplified assumption that each model only
contributes subhalo objects in the respective mass ranges. However, subhalo
abundance is a continuous function spanning all the scales $10^3$--$10^9
M_{\Sun}$ in the PDM scenario \citep{Bertschinger:2006nq}. An important
extension of our simplified picture is then to compare the PBH CDM against
standard particle DM scenarios in the full $10^3$--$10^9 M_{\Sun}$ range
taking into account the respective perturbers abundance distributions. This
will require N-body simulations down to $10^3 M_{\Sun}$ scales for both DM
models, not available to date (but see advances, e.g., from
\cite{Trashorras:2020mwn}). Note that by extrapolating particle DM results
down to the unexplored $10^3$--$10^5 M_{\Sun}$ mass range as done here
underestimates the abundance of PBH CDM clusters, and a proper computation may
increase the discriminating potential as a higher abundance of PBH objects
over shot noise would sharpen the PDF distribution. Joint classification of
other properties of the perturbers besides their mass, such as their size and
concentration, is also valuable additional information to discriminate
different DM models \citep{Bonaca:2018fek}. Besides density fluctuations
investigated here, another relevant stream property in the
line-of-parallel-angle approach is the track fluctuation \citep{Bovy:2016irg}.

Our methodologies can be readily applied to the GD-1 stream for which density
perturbations data are already available
\citep{Banik:2019cza,Hermans:2020skz}, complementing previous studies. It can
also be extended to other streams, as long as the density profile along the
stream is a good descriptor of the stellar dynamics, which is certainly the
case for other relatively cold streams such as Palomar 5 \citep{Bovy:2016irg}.
Depending on the stream, perturbations induced by baryonic objects, subleading
for our GD-1-like case, may be relevant and needed to be modeled accurately as
they may induce similar effects as DM subhalo objects of large mass. Current
observations from the GAIA and Dark Energy Survey (DES) missions are
increasingly improving data about known streams and discovering new streams
\citep{2019MNRAS.490.3508L,2020arXiv201205245I}, that will be further
complemented with upcoming observations such as DESI
\citep{2016arXiv161100036D} and the Rubin Observatory \citep{Ivezic:2008fe},
which provide new promising ground for dark matter constraints.

\section*{Acknowledgements}

We thank Alex Drlica-Wagner and Ethan O. Naddler for useful discussion. We
acknowledge use of the Hydra cluster at IFT-UAM/CSIC (Madrid). This work is
supported by the Research Project PGC2018-094773-B-C32 [MINECO-FEDER] and the
Centro de Excelencia Severo Ochoa Program SEV-2016-0597.

\section*{Data Availability}

The code underlying this article is available at
\url{https://gitlab.com/montanari/stream-dm}.

%%%%%%%%%%%%%%%%%%%%%%%%%%%%%%%%%%%%%%%%%%%%%%%%%%

%%%%%%%%%%%%%%%%%%%% REFERENCES %%%%%%%%%%%%%%%%%%

% The best way to enter references is to use BibTeX:

%%%%%%%%%%%%%%%%%%%%%%%
%% Elsevier bibliography styles
%%%%%%%%%%%%%%%%%%%%%%%
%% To change the style, put a % in front of the second line of the current style and
%% remove the % from the second line of the style you would like to use.
%%%%%%%%%%%%%%%%%%%%%%%

%% Numbered
% \bibliographystyle{model1-num-names}

%% Numbered without titles
% \bibliographystyle{model1a-num-names}

%% Harvard
% \bibliographystyle{model2-names.bst}\biboptions{authoryear}

%% Vancouver numbered
% \usepackage{numcompress}\bibliographystyle{model3-num-names}

%% Vancouver name/year
% \usepackage{numcompress}\bibliographystyle{model4-names}\biboptions{authoryear}

%% APA style
% \bibliographystyle{model5-names}\biboptions{authoryear}

%% AMA style
% \usepackage{numcompress}\bibliographystyle{model6-num-names}

%% `Elsevier LaTeX' style
\bibliographystyle{elsarticle-num}
%%%%%%%%%%%%%%%%%%%%%%%

\bibliography{biblio} % if your bibtex file is called example.bib

%%%%%%%%%%%%%%%%%%%%%%%%%%%%%%%%%%%%%%%%%%%%%%%%%%

%%%%%%%%%%%%%%%%% APPENDICES %%%%%%%%%%%%%%%%%%%%%

\appendix

\section{Convergence tests for small mass perturbers}
\label{sec:convergence}

In this section we repeat convergence tests explored in \cite{Bovy:2016irg} to
verify the reliability of numerical simulations for small masses perturbers
$10^3$--$10^5 M_{\Sun}$, and we study the effect of different perturber
internal profiles. The range $10^5$--$10^9 M_{\Sun}$ has already been
extensively explored in \cite{Bovy:2016irg}, finding overall good convergence,
except for the impact parameter factor $X$ at the largest scales. Following
closely \cite{Bovy:2016irg}, to which we refer for further details, we compute
power spectra of fluctuations in the density, $P_{Q\rho Q\rho}$, in the mean
track in parallel frequency $\langle\Delta\Omega_{\parallel}\rangle$,
$P_{\Omega\Omega}$, and their cross-spectrum, $P_{Q\rho\Omega}$. The power
spectrum is computed Fourier-transforming the density profile normalized to
the smooth one as a function of the parallel angle offset from the progenitor,
$Q\rho(\Delta\theta_{\parallel}) =
\rho(\Delta\theta_{\parallel})/\bar\rho(\Delta\theta_{\parallel})$, which
gives the dependence over the wave-number $k_\theta$. We proceed similarly for
the mean track. For each case we compute at least $1000$ realizations to
estimate the respective scatter, and show the simulations interquantile ranges
as gray bands in the following figures.

\begin{figure*}
  \centering
  \includegraphics[width=\textwidth]{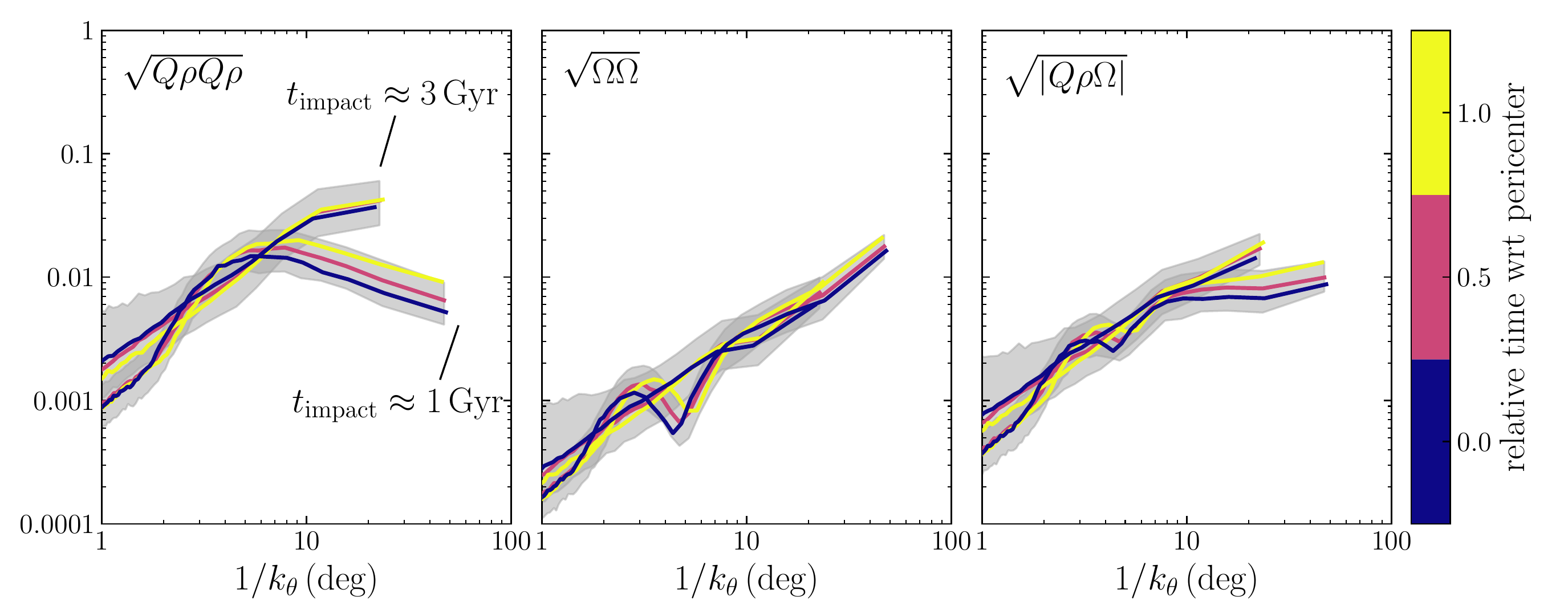}
  \caption{Power spectrum, as a function of the inverse Fourier mode, induced
    by $10^3$--$10^5\ M_{\odot}$ perturbers on a GD1-like stream if impacts
    happen at a single time, $\sim 1$~Gyr or $\sim 3$~Gyr. Different line
    colors correspond to a given orbital phase: pericenter, apocenter, or in
    between (the orbital period is $\sim 400$~Myr). Gray bands show the
    interquantile range for the set of simulations relative to a given case
    and solid curves show the median power spectrum.}
  \label{fig:conv_phase}
\end{figure*}

First, we test the importance of the sampling of the orbital phase. As
described in section~\ref{sec:sims}, we sample impacts at discrete time values
from the start of disruption to today. The sampling is chosen to be smaller
than the orbital period, but it is independent of the orbital phase. To check
whether this is a good strategy, we simulate the stream evolution assuming all
of the impacts to happen at a single time, chosen to be close to the
pericenter, apocenter or in between. We repeat this for two sets of time
values $\sim 3$~Gyr and $\sim 1$~Gyr. Results are shown in
figure~\ref{fig:conv_phase}. The two sets of impact times lead to
significantly different spectra, which is expected given that impacts at $\sim
1$~Gyr have less time to evolve than impacts at $\sim 3$~Gyr, leading to less
power. However, the dependence on the orbital phase is not statistically
relevant for our purposes. \cite{Bovy:2016irg} shows that considering larger
perturber masses the relevance of the orbital phase is even smaller. Compared
to \cite{Bovy:2016irg} results, figure~\ref{fig:conv_phase} shows smaller
power (actually smaller than the shot noise level described in
\autoref{sec:sims}) due to the smaller perturbers mass.

\begin{figure*}
  \centering
  \includegraphics[width=\textwidth]{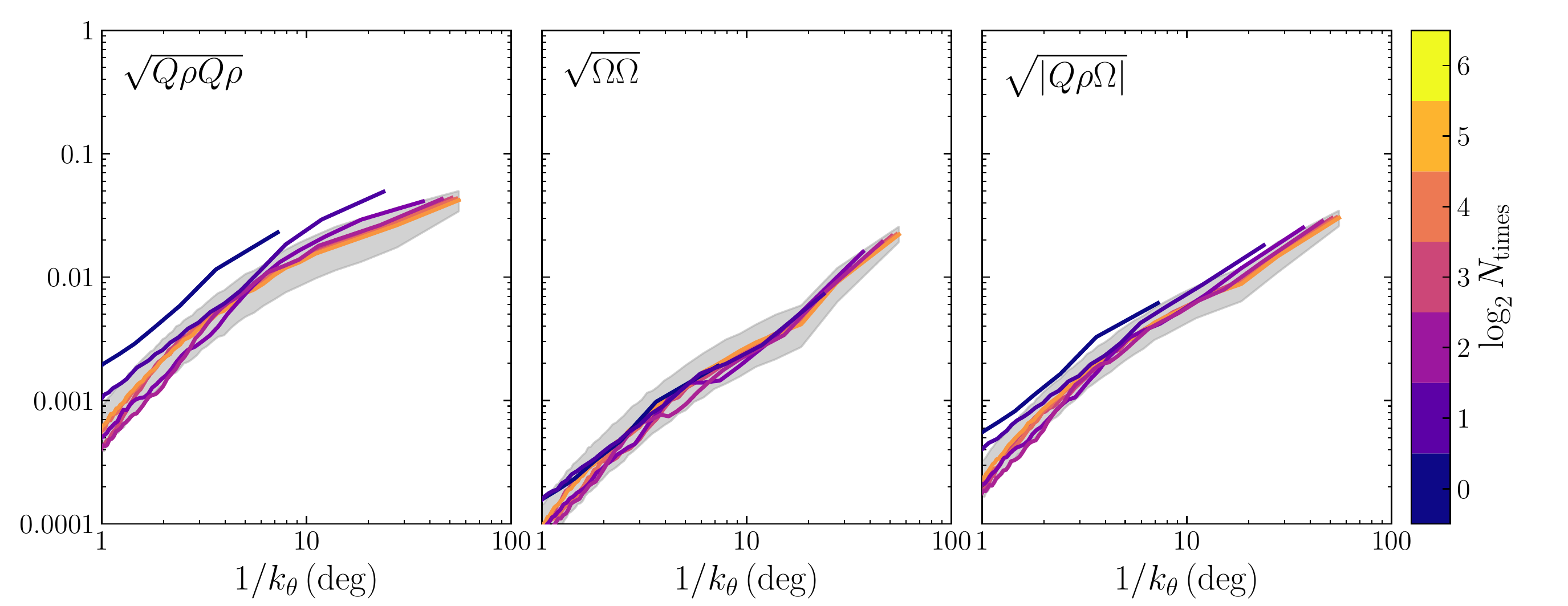}
  \caption{Power spectrum convergence for different values of the
    number of impact times since the start of the disruption of the
    stream. Results converge when the time interval between two
    impacts is comparable or smaller than the orbital period
    ($\sim 400$ Myr) of the $9$ Gyr-old stream.}
  \label{fig:conv_timesampling}
\end{figure*}

Figure~\ref{fig:conv_timesampling} shows the effect of different time
samplings. Power spectra are already well converged considering only
16 impact times, corresponding to intervals of $\sim 560$ Myr. This
suggests that a time sampling comparable to the orbital period of the
stream ($\sim400$ Myr) recovers well its statistical properties. In
the convergence tests that follow, however, we consider 32 impact
times (spaced by $\sim280$ Myr) so that the time interval between two
impacts is smaller than the orbital period. Furthermore, in the main
analysis we set an even more conservative choice (although more
computationally requiring) of 64 impact times, corresponding to
intervals of $\sim140$ Myr. The larger mass range shows similar good
convergence \citep{Bovy:2016irg}.

\begin{figure*}
  \centering
  \includegraphics[width=\textwidth]{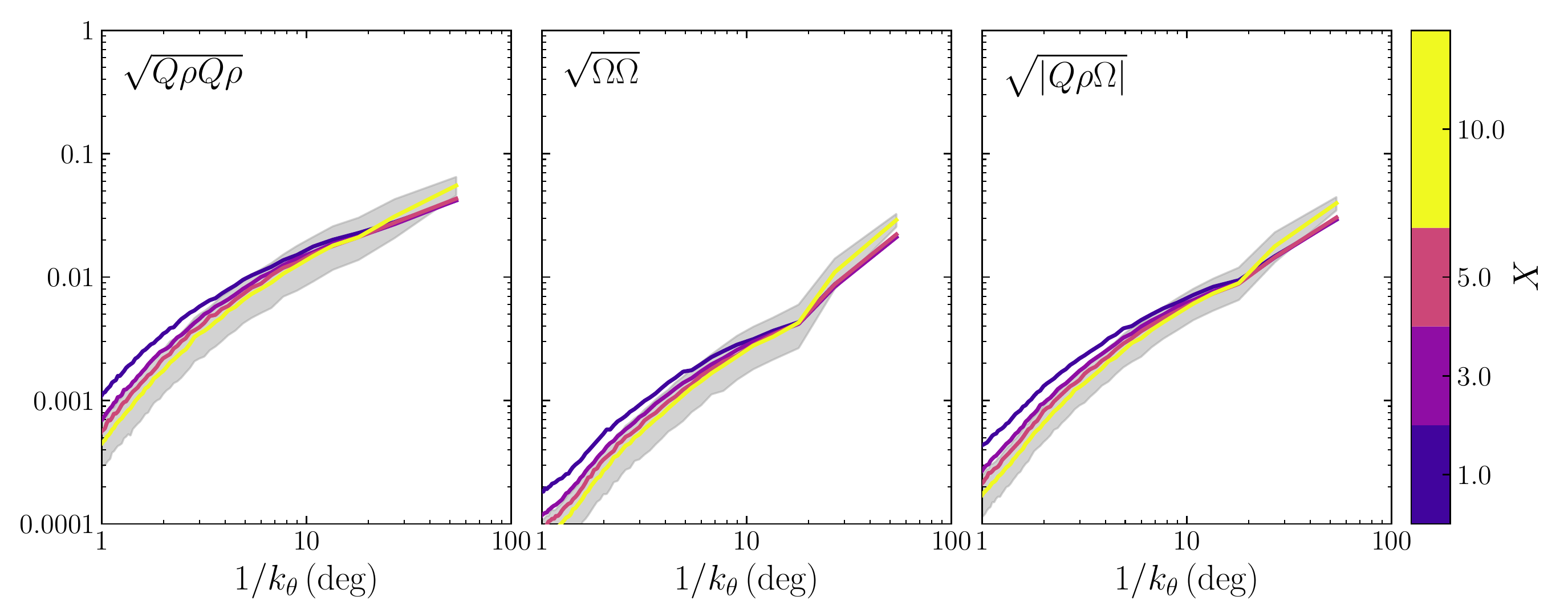}
  \caption{Power spectrum convergence for different values of the
    impact parameter factor $X$. (Note that the larger mass range
    $10^5$--$10^9 M_{\Sun}$, not shown here, do not fully converge at
    the largest scales.)}
  \label{fig:conv_bmax}
\end{figure*}

Figure~\ref{fig:conv_bmax} shows the effect of different impact
parameter factors $X$. Small scales converge for $X\gtrsim 3$ and (our
fiducial value in the main analysis and in other convergence tests is
$X=5$). However, note that the larger mass range
$10^5$--$10^9 M_{\Sun}$ do not fully converge at the largest scales
($\gtrsim 30^\circ$) as the impulse approximation is no longer
reliable and distant encounters are important \citep{Bovy:2016irg}.

\begin{figure*}
  \centering
  \includegraphics[width=\textwidth]{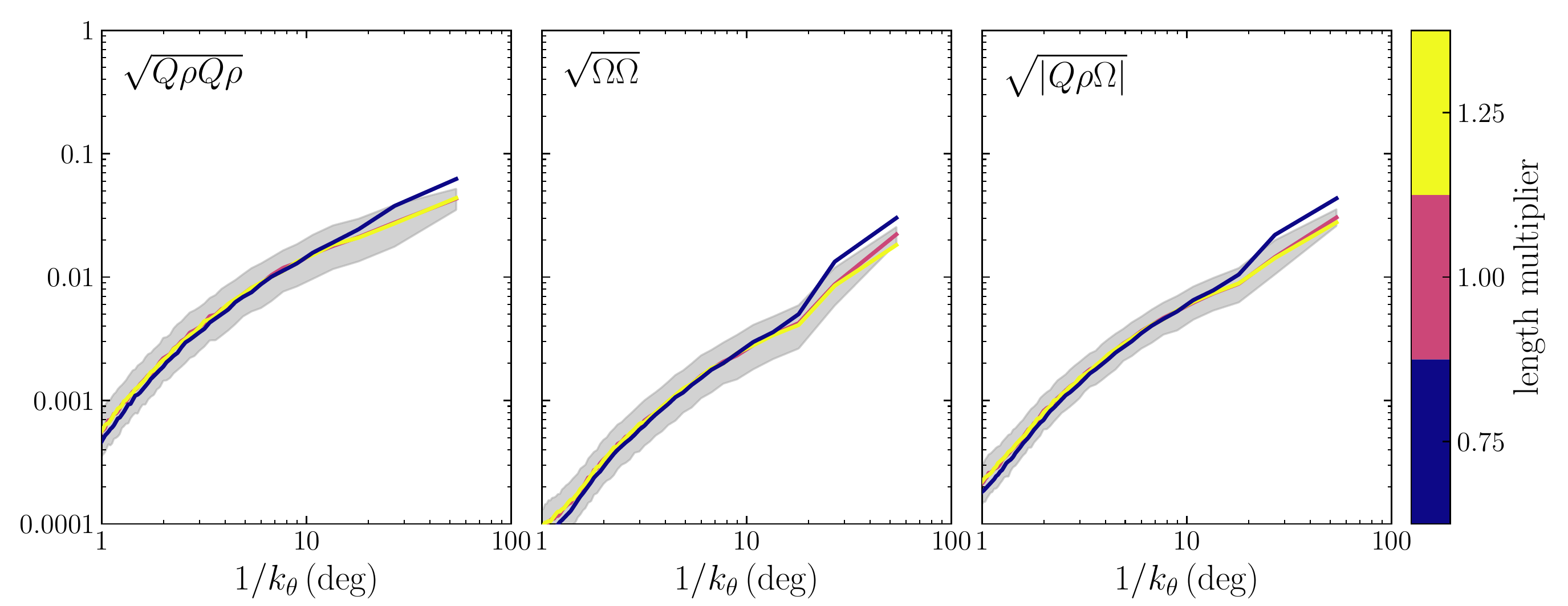}
  \caption{Power spectrum convergence for different values of the
    length scale, relative to the fiducial stream length, within which
    impacts are sampled.}
  \label{fig:conv_lenfac}
\end{figure*}

Figure~\ref{fig:conv_lenfac} shows the effect of different length factors
determining up to what distance from the progenitor impacts are sampled. The
length scale coincide with the stream length, whose edge is defined as the
location along the stream where the density drops by 20\% compared to the
respective value close to the progenitor. Setting different length multipliers
to sample impacts up to 75\% and 125\% of the fiducial length shows good
convergence (to keep the same number of impacts for all the cases we also
scale the predicted density of perturbers together with the length factor).
The larger mass range shows similar good convergence \citep{Bovy:2016irg}.

\begin{figure*}
  \centering
  \includegraphics[width=\textwidth]{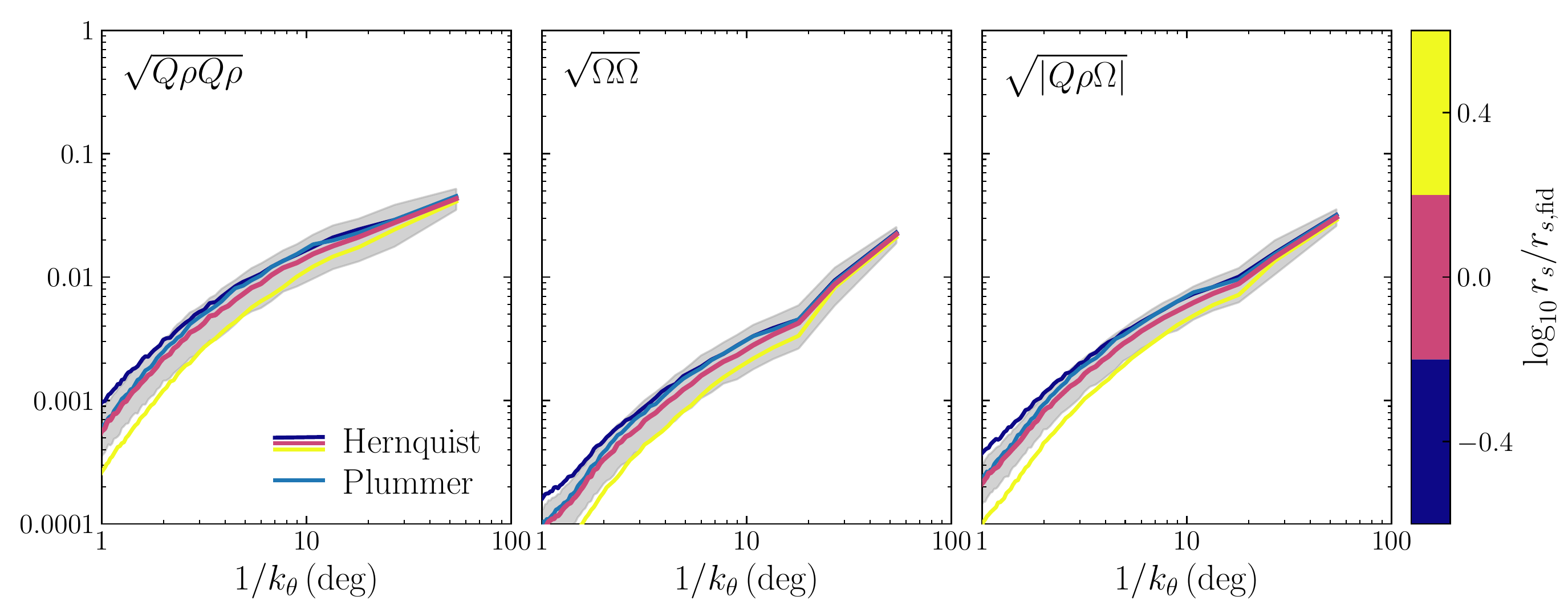}
  \caption{Power spectrum for different scaling of the fiducial
    Hernquist scale parameter $r_s(M)$, and for a plummer profile.}
  \label{fig:profile}
\end{figure*}

Figure~\ref{fig:profile} compares different perturber profiles. The
fiducial Hernquist profile scale is set by
$r_{s, {\rm fid}}(M) = 1.05 {\rm kpc}\ (M / 10^8 M_\odot)^{0.5}$,
obtained by fitting the Via Lactea II simulation
\citep{Bovy:2016irg,2008Natur.454..735D}. We first investigate a
constant factor scaling 2.5 times larger or smaller than the fiducial
relation (for consistency with the maximum sampled value of the impact
parameter, we also set an impact parameter factor $X$ such that
$X = X_{\rm fid} r_{s, {\rm fid}} / r_s$, where $X_{\rm fid}=5$),
which roughly corresponds to the scatter observed in the Via Lactea II
simulation. Then we also compare the power spectrum obtained assuming
a Plummer profile $r_s(M) = 1.62 {\rm kpc}\ (M / 10^8 M_\odot)^{0.5}$
(also here for consistency we need to set an impact parameter factor
$X$ such that $X=(1.05/1.62) X_{fid}$, where we take the ratio of the
constant factors appearing in the Hernquist and Plummer scale
parameters). Although more compact perturbers induce larger power on
the smallest scales, differences are negligible for our purposes, as
is the case for the larger mass range \cite{Bovy:2016irg}. Deviations
are particularly small when comparing Hernquist and Plummer profiles,
the main difference being the inner region of the perturber (cuspy for
the first case, and smoothed in the second one) that is not relevant
for our purposes.

\section{Power spectrum dependence on mass range}
\label{sec:powspec}

\begin{figure}
  \centering
  \includegraphics[width=0.7\textwidth]{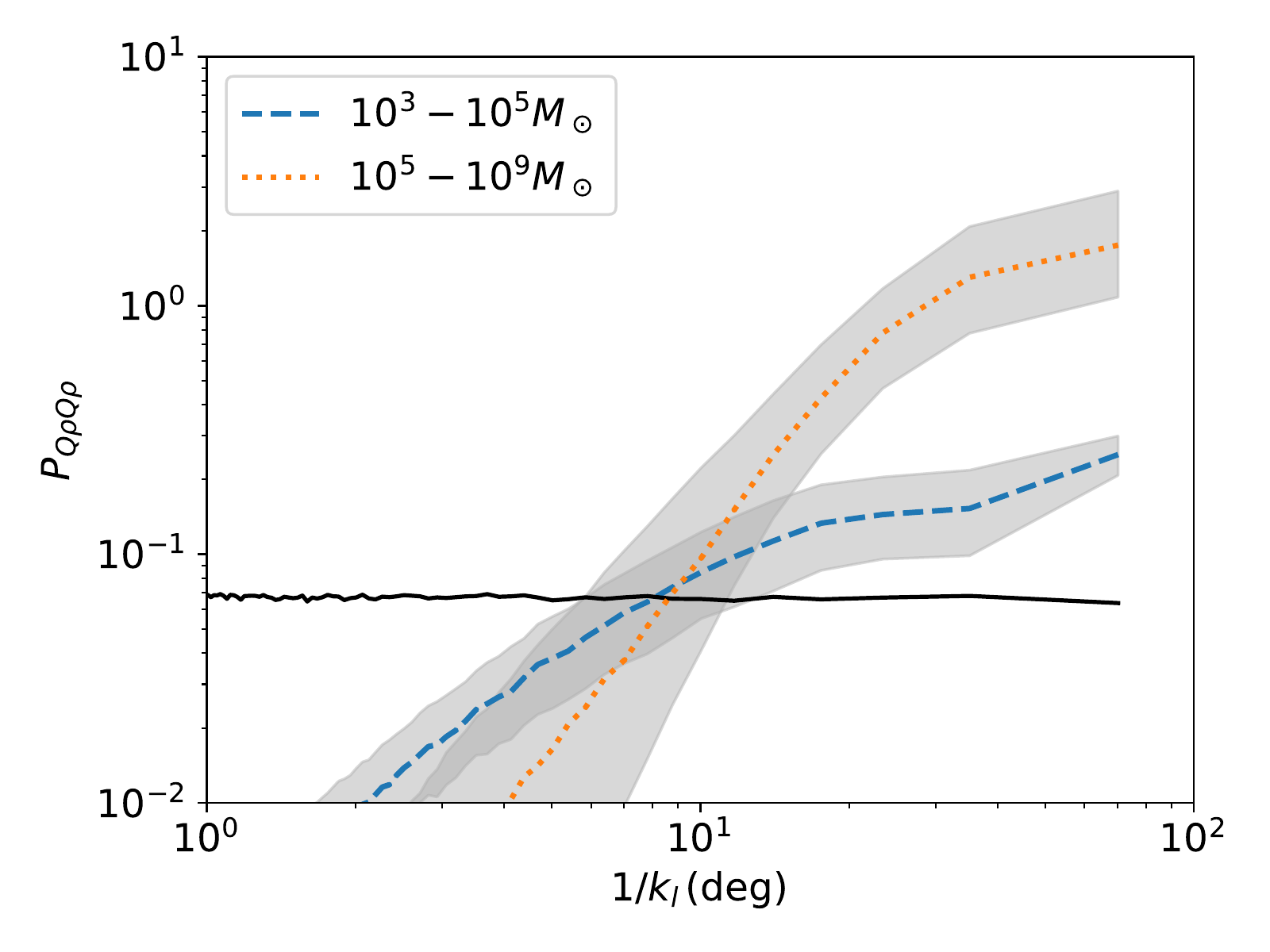}
  \caption{Power spectrum of density fluctuations $Q\rho$ for the mass
      ranges considered in the main analysis. Gray bands show the
      interquantile range. The solid line shows the shot noise contribution.}
  \label{fig:powspec}
\end{figure}

In this section we show the power spectrum of density fluctuations $Q\rho$ for
both mass ranges considered in the main analysis. The computation and
interpretation of the power spectrum proceeds similarly to
\ref{sec:convergence}, but we Fourier-transform the density as a
function of the observable Galactic longitude $Q\rho(l)$, which gives the
dependence over the wave-number $k_l$. This allows us to include the
contribution of shot noise (see \autoref{sec:sims}) as the median of $10^3$
realizations of Gaussian noise.

Both cases are detectable above shot noise over $1/k_l > 8$~deg scales and
they are clearly distinguishable at the largest scales, over $1/k_l > 20$~deg.
While it would be interesting to compare model selection performance based on
these power spectra to the methods outlined in the main analysis, a proper
estimate of the Bayesian evidence is computationally prohibitive for the power
spectrum. Note that our PDF methodology provides robust results for the Bayes
factors relying on only $\mathcal{O}(10^3)$ simulations thanks to the fact
that we disregard bins where the evidence computation is not reliable due to
poor sampling, see \autoref{sec:res_pdf}. While the power spectrum has the
advantage over the PDF of retaining spatial information, density fluctuations
are highly non-Gaussian so the power spectrum is an incomplete summary
statistics and the infinite series of higher correlation moments is expected
to contain valuable complementary information. Neither of the gradient
boosting nor the PDF method proposed in the main analysis suffer from this
limitation.

%%%%%%%%%%%%%%%%%%%%%%%%%%%%%%%%%%%%%%%%%%%%%%%%%%

\end{document}